\documentclass[aps,prd,superscriptaddress,twocolumn,,amssymb,eqsecnum,showpacs,showkeyes,secnumarabic,graphics,floatfix,nofootinbib,tightenlines,longbibliography]{revtex4-1}
\usepackage{float}
\usepackage{graphicx}
\usepackage{epsfig}
\usepackage{bm}
\usepackage{cancel}
\usepackage{amsmath}
\usepackage[toc,page]{appendix}
\usepackage[dvipsnames]{xcolor}
\usepackage[margin=3cm]{geometry}
\usepackage{hhline}
\usepackage[utf8]{inputenc}
\usepackage[breaklinks=true,colorlinks=true,
linkcolor=blue,urlcolor=blue,citecolor=cyan,
bookmarks=true,bookmarksopenlevel=2]{hyperref}

\def \beq  {\begin{equation}}
\def \eeq  {\end{equation}}
\def \ber  {\begin{eqnarray}}
\def \eer  {\end{eqnarray}}

\begin{document}
\newcommand{\newc}{\newcommand}

\newc{\be}{\begin{equation}}
\newc{\ee}{\end{equation}}
\newc{\ba}{\begin{eqnarray}}
\newc{\ea}{\end{eqnarray}}
\newc{\bea}{\begin{eqnarray*}}
\newc{\eea}{\end{eqnarray*}}
\newc{\ie}{{\it i.e.} }
\newc{\eg}{{\it e.g.} }
\newc{\etc}{{\it etc.} }
\newc{\etal}{{\it et al.}}
\newc{\lcdm}{$\Lambda$CDM}
\newcommand{\nn}{\nonumber}

\date{\today}
\title{\bf Spinning particle orbits around a black hole in an expanding background}

\author{I. Antoniou}\email{ianton@uoi.gr}
\affiliation{Department of Physics, University of Ioannina,
GR-45110, Ioannina, Greece}
\author{D. Papadopoulos}\email{papadop@astro.auth.gr}
\affiliation{Department of Physics, University of Thessaloniki,
Thessaloniki,54124, Greece}
\author{L. Perivolaropoulos}\email{leandros@uoi.gr}
\affiliation{Department of Physics, University of Ioannina,
GR-45110, Ioannina, Greece}

\begin{abstract}
We investigate analytically and numerically the orbits of spinning particles around black holes in the post Newtonian limit and in the presence of cosmic expansion. We show that orbits that are circular in the absence of spin, get deformed when the orbiting particle has spin. We show that the origin of this deformation is twofold: a. the background expansion rate which induces an attractive (repulsive) interaction due to the cosmic background fluid when the expansion is decelerating (accelerating) and b. a spin-orbit interaction which can be attractive or repulsive depending on the relative orientation between spin and orbital angular momentum and on the expansion rate.
\end{abstract}
\pacs{04.20.Cv, 97.60.Lf, 98.62.Dm}
\maketitle

\begin{section}{Introduction}
Even though most astrophysical bodies have spins and evolve in an expanding cosmological background, their motion is described well by ignoring the cosmic expansion and under the nonspinning test particle approximation for large distances from a central massive body and for relatively low spin values \cite{Hartl:2002ig}. These approximations however become less accurate for large values of the spin and/or when the mass of the cosmic fluid inside the particle orbit becomes comparable to the mass of the central massive object. For such systems new types of interactions appear which are proportional to the time derivatives of the cosmic scale factor and the spin of the orbiting particle. For example, phantom dark energy models can lead to dissociation of all bound systems in the context of a Big-Rip future singularity \cite{Caldwell:2003vq,Nesseris:2004uj,Antoniou:2016obw,Nolan:2014maa}. Also, the spin-curvature interaction \cite{Obukhov:2013zca} can modify the motion of the test particles in black hole spacetimes \cite{Tod:1976ud,Semerak:1999qc,Khriplovich:2008ni,Lukes-Gerakopoulos:2016udm,Costa:2012cy} due to spin-spin or spin-orbit couplings \cite{Burko:2003rv,Shibata:1993uk,Han:2016djt}, or make the motion chaotic \cite{Suzuki:1999si,Kubiznak:2011ay,Kao:2004qs} thus modifying significantly the orbit of the test body leading to the emission of characteristic forms of gravitational waves \cite{Saijo:1998mn,Tanaka:1996ht,Mino:1995fm,Harms:2016ctx}.

Such interactions have been investigated previously for nonspinning test particles in an expanding background around a massive body (McVittie background \cite{Kaloper:2010ec}) and it was shown that accelerating cosmic expansion can lead to dissociation of bound systems in the presence of phantom dark energy with equation of state parameter $w<-1$ \cite{Caldwell:2003vq,Nesseris:2004uj,Antoniou:2016obw}. In the absence of expansion but in the presence of spin for the test particles it has been shown that spin-orbit and spin-spin interactions in a Kerr spacetime  can lead to deformations of circular orbits for large spin values \cite{Han:2016djt}. In view of these facts, the following interesting questions emerge
\begin{enumerate}
\item
Are there circular orbit deformations for spinning test particles embedded  in the post Newtonian limit of McVittie background (Schwarzschild metric embedded in an expanding background)? Such deformations could be anticipated due to the coupling of the particle spin with its orbital angular momentum.
\item
What is the nature of such deformation and how do the corresponding deformations depend on the orientation of the spin with respect to the angular momentum?
\item
How do these deformations depend on the nature of the background expansion?
\end{enumerate}
These questions are addressed in the present analysis.

The structure of this paper is the following: In the next section we briefly review the Mathisson-Papapetrou (MP) equations \cite{Mathisson:1937zz} and the common supplementary conditions, we introduce the McVittie background corresponding to a black hole embedded in an expanding background and its post Newtonian limit. In section III we discuss the conserved quantities of a spinning test particle in a given spherically symmetric  metric in an expanded background, we consider the post Newtonian limit of McVittie metric and construct the geodesic equations of a spinning particle using the Mathisson-Papapetrou equations. We also solve these equations numerically and identify the deformation of orbits due to the presence of test particle spin. We identify the dependence of this deformation on the relative orientation between spin and orbital angular momentum of the spinning test particle. Finally, in section IV we summarize, discuss the implications of our results and identify possible future extensions of our analysis.

\end{section}

\begin{section}{The Equations of Motion of a Spinning Particle. The MP Equations.}

Consider a massive spinning test particle, in MP’s model \cite{Papapetrou:1951pa, Mathisson:1937zz}. The equations of motion of a spinning particle originally derived from Papapetrou ($1951$) and later on reformulated by Dixon \cite{Dixon1964,Dixon:1970zza} can be extracted through the corresponding Hamiltonian \cite{Barausse:2009aa,Kunst:2015tla} or through the extremization of the corresponding action \cite{Cho:1997vx}, whose variation is \cite{Porto:2005ac} 
\begin{equation}
    \delta L=-p^\mu \delta \upsilon_\mu-\frac{1}{2}S^{\mu\nu}\delta\Omega_{\mu\nu}\label{lag}
\end{equation}
where $\upsilon^\mu=\frac{dx^\mu}{d\tau}$ is the four-velocity of the test particle tangent to the orbit $x^\mu = x^\mu(\tau)$, $\tau$ is the proper time across the worldline $x^\mu(\tau)$, $p^\mu$ is its four-momentum and $S^{\mu \nu}$ are the components of the antisymmetric spin tensor. Also, $\Omega_{\mu\nu}=\eta^{IJ}e_{\mu I}\frac{De_{\nu J}}{d\tau}$ is an antisymmetric tensor, $\eta^{IJ}=e^\mu_Ie^\nu_Jg_{\mu\nu}$ and $e^\mu_I$ is a tetrad attached to each point of the worldline. 

The MP equations are of the form \cite{Porto:2005ac,Apostolatos, Chicone:2005jj,Steinhoff:2010zz}:
\be \frac{Dp^\mu}{d\tau} \equiv \frac{dp^\mu}{d\tau}+\Gamma^\mu_{\lambda \nu}\upsilon^\lambda p^\nu=-\frac{1}{2}R^\mu_{\nu \lambda \rho}S^{\lambda \rho}\upsilon^\nu\label{e2}\ee
\begin{equation} 
\begin{aligned}
 \frac{DS^{\mu \nu}}{d\tau} &\equiv \frac{dS^{\mu \nu}}{d\tau}+ \Gamma^\mu_{\lambda \rho}\upsilon^\lambda S^{\rho \nu}+\Gamma^\nu_{\lambda \rho}\upsilon^\lambda S^{\mu \rho} \\ &=p^\mu \upsilon^\nu-p^\nu \upsilon^\mu \label{e3}
 \end{aligned} 
 \end{equation}
The dynamical equations imply, spin-orbit coupling, i.e., spin couples to the velocity of the orbiting spinning particle, thus deforming the geodesic. Therefore the spin force deforms the geodesic.

The spin tensor keeps track of the intrinsic angular momentum associated with a spinning particle. The term in the r.h.s. of Eq. (\ref{e2}) shows an interaction between the curvature of the spacetime and the spin of the particle. Due to the coupling between curvature and spin, the four-momentum is not always parallel to the $\upsilon^\mu$. 
This may be seen by multiplying Eq. (\ref{e3}) with $\upsilon_\nu$. Then, leads to \be p^\mu=m\upsilon^\mu-\upsilon_\nu \frac{DS^{\mu \nu}}{d\tau}\label{e4}\ee
where $m =-p^\mu \upsilon_\mu$ is the rest mass of the particle with respect to $\upsilon_\mu$.

Since $\tau$ is the proper time, the condition $\upsilon_\mu \upsilon^\mu = -1$ applies. The measure of the four-momentum \begin{equation}p_\mu p^\mu =-\mu^2\label{mu}\end{equation} provides the `total' or `effective' \cite{Semerak:1999qc} rest mass $\mu$ ($p^\mu=\mu u^\mu$) with respect to $p^\mu$ where $u^\mu$ is the `dynamical four-velocity'.  and is equal to $m$, only if $\upsilon^\mu$ coincides with the four-velocity $u^\mu$ ($u^\mu=\upsilon^\mu$). In the linear approximation of the spin $p^\mu$ and $\upsilon^\mu$ are parallel. Generally, since $u^\mu\neq\upsilon^\mu$ which means that $\frac{DS^{\mu\nu}}{d\tau}\neq 0$ (see Eq. (\ref{e4})) a spinning particle does not
follow the geodesics of the spacetime (the r.h.s. of Eq. (\ref{e2}) is non zero, since $S^{\mu\nu}\neq 0$). Therefore its motion is generalized on a world line rather than geodesics.

In the context of the MP equations the multipole moments of the particle higher than a spin dipole are ignored \cite{Dixon1973}. This is the spin-dipole approximation, because the particle is described as a mass monopole and spin dipole \cite{Mashhoon:2006fj}. The equations in quadratic order of spin have also been derived \cite{Vines:2016unv}. The MP equations can also get generalized in order to describe a test spinning particle in Modified theories of Gravity \cite{Roshan:2012qy}.

The MP equations (\ref{e2}) and (\ref{e3}) have been discussed by many authors and solutions have been presented. These solutions refer mainly to Schwarzschild background spacetime \cite{Harms:2016ctx,Plyatsko:2008rd,Plyatsko:2008wh,Plyatsko:2005bh,Plyatsko:2016bee, Suzuki:1996gm,0264-9381-10-3-017,Bini:2005nt}, to Kerr spacetime \cite{Harms:2016ctx,Plyatsko:2011gf,Plyatsko:2013xza,Lukes-Gerakopoulos:2016bup,Bini:2006pc,Hartl:2003da,Suzuki:1997by,Han:2008zzf,Hackmann:2014tga}, to de Sitter spacetime \cite{Mohseni:2010rm,Mortazavimanesh:2009rm,Obukhov:2010kn,Stuchlik:2003dt} and to FRW spacetime \cite{Zalaquett:2014eia} for chargeless or charged test spinning particles \cite{Hojman:1976kn,Bini:2000vv}. The evolution of spinning particles in spacetimes with torsion has also been investigated \cite{Nomura:1991yx,Maity:2004yk}. 

Eqs. (\ref{e2}) and (\ref{e3}) are the equations of motion for a spinning body which reduce to the familiar geodesic equations when the spin tensor $S^{\mu\nu}$ vanishes. However, they do not form a complete set of equations and we need further equations to close the system \cite{Lukes-Gerakopoulos:2017cru}. The problem of the unclosed set
of equations in (\ref{e2}) and (\ref{e3}) can be physically understood by the requirement that the particle must have a finite size which does not make the choice of the reference worldline uniquely defined \footnote{\url{https://d-nb.info/1098374932/34}}. The additional conditions used are the spin supplementary conditions (SSC) \cite{Costa:2011zn}.  When we choose a SSC, we define the evolution of the test body in a unique worldline $x^\mu(\tau)$ and we fix the center of mass (corresponds to the centre where the mass dipole vanishes), which is usually called centroid. The centroid is a single reference point inside the body, with respect to which the spin is measured \cite{Lukes-Gerakopoulos:2017vkj}.

There are several SSC but two of them are more commonly used 
\begin{itemize}
    \item The P condition (Mathisson-Pirani) \cite{Pirani:1956tn} 
\be \upsilon_\mu S^{\mu \nu} = 0 \label{e5a}\ee 
so that the spin four-vector is perpendicular to the four-velocity and implies that $\frac{d\mu}{d\tau}=0$ \cite{Lukes-Gerakopoulos:2014dma}. It does not provides a unique choice of representative worldline, as it is dependent on the observer’s velocity and therewith on the initial conditions. It is often referred to as the proper centre of mass \cite{Costa:2011zn}.
    \item The T condition (Tulczyjew-Dixon) \cite{Tulczyjew}
\be p_\mu S^{\mu \nu} = 0\label{e5b}\ee
so that the spin four-vector is perpendicular to the four-momentum and implies that $\frac{dm}{d\tau}=0$ \cite{Lukes-Gerakopoulos:2017cru}. This condition is physically correct, since the trajectory of the extended body is determined by the position of the center of mass of the body itself \cite{Bini:2014poa}. This constraint is a consequence of the theory, i.e., the Tulczyjew constraint can derived from the Lagrangian theory \cite{Asenjo:2016uxz} and restricts the spin tensor to generate rotations only.
\end{itemize}

Analytic discussions and  thorough  reviews on different choices about the SSCs may be found in refs \cite{Kyrian:2007zz, Costa:2014nta, Costa:2017kdr}. Generally, different SSC are not equivalent since every SSC defines a different centroid for the system. The author of ref. \cite{Hartl:2002ig} point out that the difference between the two conditions (\ref{e5a}) and (\ref{e5b}) is third order in the spin, so results for physically realistic spin values, are unaffected. In what follows we use the  T condition, which defines the centre of mass of the particle in the rest frame of the central gravitating body. 

The McVittie metric describes a expanding cosmological background with strong gravity, such as a spacetime near a black hole or a neutron star. In a $(t,r,\theta,\phi)$ coordinate system McVittie \cite{McVittie:1933zz} found a solution given by the equation (see  eq. (29) of ref. \cite{McVittie:1933zz} with $G=c=1$)
\begin{eqnarray}
&&ds^2=-(1-\frac{m(t)}{2r})^2(1+\frac{m(t)}{2r})^{-2}dt^2+\nonumber\\&&+(1+\frac{m(t)}{2r})^4 a^2(t)\big(dr^2+r^2d\Omega^2\big)\label{s1}
\end{eqnarray}
where $d\Omega^2=d\theta^2+\sin^2{\theta}d\phi^2$. 

The component $G^t_r$ of the Einstein tensor is \begin{equation}
G^t_r=\frac{8(2r+m)}{a(2r-m)^3}(\dot am+a\dot m)\label{gtr}
\end{equation} 
Imposing the "no-accretion" condition $G^t_r=0$ (there is no flux of relativistic mass across the equatorial surface \cite{McVittie:1933zz}) we find that $\frac{\dot{a}}{ a}=-\frac{\dot{m}}{m}$ or $m=\frac{m_0}{a(t)}$, where $m_0$ is a constant of integration and is identified with the mass of the central body  at the origin \cite{Carrera:2009ve}. The curvature of space is here assumed to be asymptotically zero.

At any instant of time $t_1$ the observer's coordinate for measuring distance from the origin is $\tilde{r}=r a(t_1)$. If we write $M=m(t_1) a(t_1)$, the metric (\ref{s1}) becomes
\begin{equation}\label{s2}
ds^2=-(\frac{1-\frac{M}{2 \tilde{r}}}{1+\frac{M}{2\tilde{r}}})^2 dt^2+(1+\frac{M}{2\tilde{r}})^4 \big(d\tilde{r}^2+\tilde{r}^2d\Omega^2\big)
\end{equation}

In the weak field limit we have $\frac{M}{2r}\ll1$, ie

\begin{equation}
ds^2=-(1-\frac{2GM}{r})dt^2+(1+\frac{2GM}{r})(dr^2+r^2d\Omega^2)\label{mc1}
\end{equation}
which is the \emph{Newtonian limit} of Schwarzschild's spacetime.

Setting $ r=a(t)\rho$ and $R_s=2M$ the metric (\ref{mc1}) reads

\be ds^2=-(1-\frac{R_s}{a\rho})dt^2+a^2(1+\frac{R_s}{a\rho})(d\rho^2+\rho^2d\Omega^2)\label{mcvittief}\ee

For a static background ($a = 1$) the metric (\ref{mcvittief}) becomes the Schwarzschild metric in isotropic coordinates (the spacelike slices are as close as possible to Euclidean) as expected \cite{schutz_2009}, while for $R_s=0$ becomes the FRW metric in spherical coordinates.

The `areal' radius \cite{Carrera:2006im} of the metric (\ref{mcvittief}) is equal to the square root of the modulus of the coefficient of the angular part $d\Omega^2$ of the metric, namely \begin{equation}
    R(t,\rho)=(1+\frac{R_s}{a\rho})^{1/2}a\rho\label{areal}
\end{equation}
and the corresponding modulus of angular momentum, which is a constant of motion for a spinless particle, defined as \begin{equation}
   \mathcal{L}=R^2(t,\rho)\dot \phi\label{l}
\end{equation}
\end{section}

\begin{section}{Spinning particle in McVittie spacetime-Post Newtonian Limit} 
\subsection{The MP equations in an expanding Universe}
 
We consider the case where the spinning particle orbits on the equatorial plane, which means that $\theta=\pi/2$. Also, on the equatorial plane valid $\upsilon^2\equiv\upsilon^{\theta}=0$ and $p^\theta=0$ since $p^{\mu}=\frac{\mu^2}{m}\upsilon^{\mu}$. The metric (\ref{mcvittief}) is independent of the $\phi$ coordinate, therefore admits a $\phi$-Killing vector e.g. $\xi^{\mu}=(0,0,0,1)$ which gives
\begin{equation}\label{f1}
J_z=p_\mu \xi^{\mu}-\frac{1}{2}\xi_{\mu,\nu}S^{\mu\nu}
\end{equation} or \begin{equation}
J_z=p_\phi-\frac{1}{2}g_{\phi\mu,\nu}S^{\mu\nu}\label{f1a}
\end{equation} where $J_z$ is the $z$ component of the angular momentum, which is a conserved quantity of the motion of a spinning particle. This constant of motion exists independently of the choice of the supplementary condition and reflects the symmetry of the background spacetime. 

The spin tensor has six independent components but since we demand equatorial planar motion, the particle must have angular momentum only in $z$ axis ($J_z\neq 0$). The conditions $J_x=0$, $J_y=0$ and $p^\theta=0$ (necessary conditions for motion in the equatorial plane) require that $S^{r\theta}=0$ and $S^{\theta\phi}=0$. Also, the absence of acceleration perpendicular to the equatorial plane implies that $S^{t\theta}=0$ \cite{Zalaquett:2014eia}. Thus, planar motion requires alignment of the spin with the orbital angular momentum and the motion characterized only by three independent spin components. With these assumptions, the spin tensor becomes a vector and the formulation will be simpler. From the T condition (\ref{e5b}) we derive the spin components $S^{03}$ and $S^{13}$ in terms of $S^{01}$ as
\begin{eqnarray}\label{n1a}
S^{03}&=&-\frac{p_1}{p_3}S^{01},\nonumber\\
S^{13}&=&\frac{p_0}{p_3}S^{01}
\end{eqnarray}

In order to complete the system of Eqs. (\ref{e2}) and (\ref{e3}) we have to add two more equations, corresponding to conserved quantities in the context of the T condition. The first is the dynamical mass $\mu$ \cite{Hojman:2012me} with respect to the four-momentum $p^\mu$ which defined through Eq. (\ref{mu})
and the second is the particle's total spin $s$ which is defined as the positive root of \begin{equation}\label{n8}
s^2=\frac{1}{2}S_{\mu\nu}S^{\mu\nu}
\end{equation} The first derivative of $s^2$ with respect to $\tau$ is $\dot{s^2}=2p_\mu S^{\mu\nu}\upsilon_\nu$ \cite{Lukes-Gerakopoulos:2017cru} which vanishes in the context of T condition.
From (\ref{n8}) we have
 \begin{equation}\label{n8c}
 s^2=\frac{(1-\xi)(S^{01})^2}{\rho^2 (p^3)^2}\mu^2
\end{equation}
where 
\begin{equation}\label{n8aa}
\xi\equiv\frac{R_s}{a\rho}\ll1
\end{equation}

Using the Eqs. (\ref{mu}) and (\ref{n8c}) we define the parameter $\Omega^2$ as the ratio
\begin{equation}\label{n8d}
\Omega^2\equiv\frac{s^2}{\mu^2}=\frac{(S^{01})^2}{\rho^2 (p^3)^2}(1-\xi)
\end{equation} which is a constant of motion, since $\mu$ and $s$ are conserved quantities.
From Eq. (\ref{n8d}) it is easy to calculate the spin component $S^{01}$ 
\begin{equation}\label{n8e}
\frac{S^{01}}{p^3}=\frac{\rho \Omega}{\sqrt{1-\xi}}
\end{equation}

Thus, from Eqs. (\ref{n1a}) and (\ref{n8e}) the non zero spin components in our consideration are 
\begin{eqnarray}\label{n1}
S^{01}&=&\frac{\rho \Omega p^3}{\sqrt{1-\xi}}\nonumber\\
S^{03}&=&-\frac{p^1}{\rho^2p^3}S^{01},\nonumber\\
S^{13}&=&-\frac{(1-2\xi)p^0}{a^2\rho^2p^3}S^{01}
\end{eqnarray}

Using now the post Newtonian limit of McVittie metric (\ref{mcvittief}), starting from the MP equation (\ref{e2}) and setting the index $\mu=1$ it is straightforward  to derive the radial geodesic equation for the spinning particle. We replace the distance $\rho$ as $\rho=r/a$ and the corresponding derivatives with respect to $t$, $\dot\rho=d\rho/dt$ and $\ddot\rho=d^2\rho/dt^2$. Also, we ignore terms of order $(R_s)^2$ (post Newtonian limit) and the final result is
\begin{equation}
\begin{aligned}
&\ddot{r}-\frac{\ddot{a}}{a}r-r\dot{\phi}^2=-r\Omega\dot\phi\bigg(\frac{\ddot a}{a}-\frac{\dot a^2}{a^2}\bigg)+\\ &+\frac{R_s}{2}\bigg(-\frac{1}{r^2}-\dot\phi^2-\frac{\dot a^2}{a^2}+\frac{\dot r^2}{r^2}-\frac{3\Omega\dot\phi}{r^2}\bigg) \label{mvr}
\end{aligned}
\end{equation}
Similarly, from the MP equation (\ref{e2}) and setting the index $\mu=3=\phi$ we obtain
\begin{equation}
\frac{d\big(r^2\dot{\phi}\big)}{dt}=\dot r\dot\phi R_s+\Omega r^2\bigg(\frac{\dot r}{r}-\frac{\dot a}{a}\bigg)\bigg(\frac{\ddot a}{a}-\frac{\dot a^2}{a^2}\bigg)\label{mvphi}
\end{equation}
which would lead to orbital angular momentum conservation in the absence of spin ($\Omega=0$). Indeed, the first derivative of Eq. (\ref{l}) with respect to time must be zero and gives the Eq. (\ref{mvphi}) for a spinless particle \cite{Carrera:2006im}.

\begin{figure*}[ht]
\centering
\begin{center}
$\begin{array}{@{\hspace{0.0in}}c@{\hspace{0.0in}}c@{\hspace{0.0in}}c}
\multicolumn{1}{l}{\mbox{}} 
\epsfxsize=2.98in
\epsffile{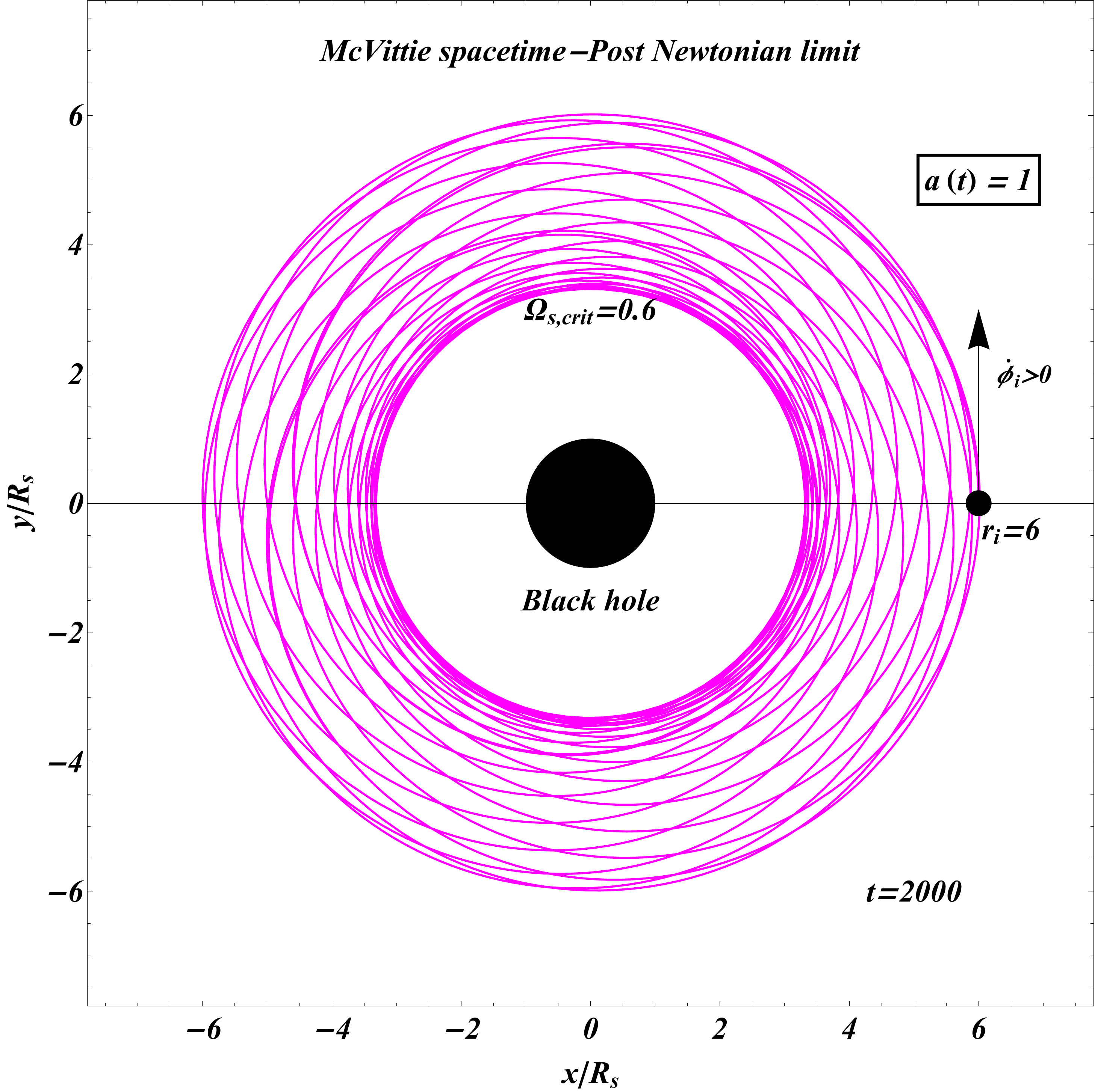} &
\epsfxsize=3in
\epsffile{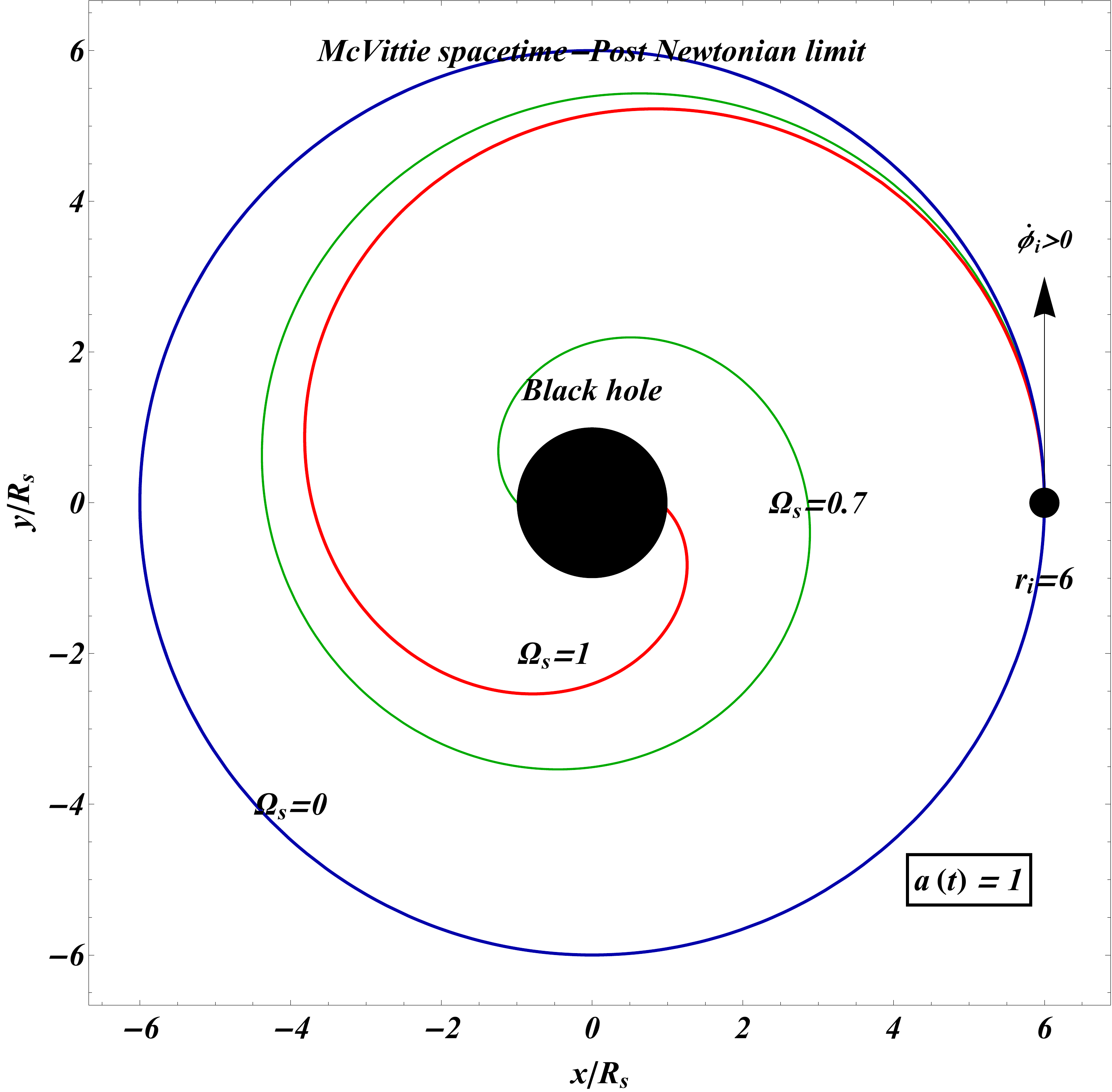} 
\end{array}$
\end{center}
\vspace{0.0cm}
\caption{Spinning particle orbits in a static universe. The circular orbits that would be present for a non-spinning particle get disrupted due to the spin-orbit coupling in the presence of spin. For $ \Omega_s\dot{\phi}>0$ the spin-orbit coupling force is attractive and the circular orbits are deformed inward. The left panel (where $\Omega_s=0.6$) corresponds to maximum (critical) value of $\Omega_s$, for which the particle remains bounded. The innermost stable circular orbit (ISCO) is $3R_s$. When $\Omega_s>0.6$, at some time the radius of the orbit becomes less than $3R_s$ and the particle is captured by the black hole (right panel). For non-spinning particle ($\Omega_s=0$) the circular orbits shown in right panel remain undisrupted.}
\label{fig1}
\end{figure*} 

\begin{figure*}[ht]
\centering
\begin{center}
$\begin{array}{@{\hspace{0.0in}}c@{\hspace{0.0in}}c@{\hspace{0.0in}}c}
\multicolumn{1}{l}{\mbox{}} 
\epsfxsize=3.1in
\epsffile{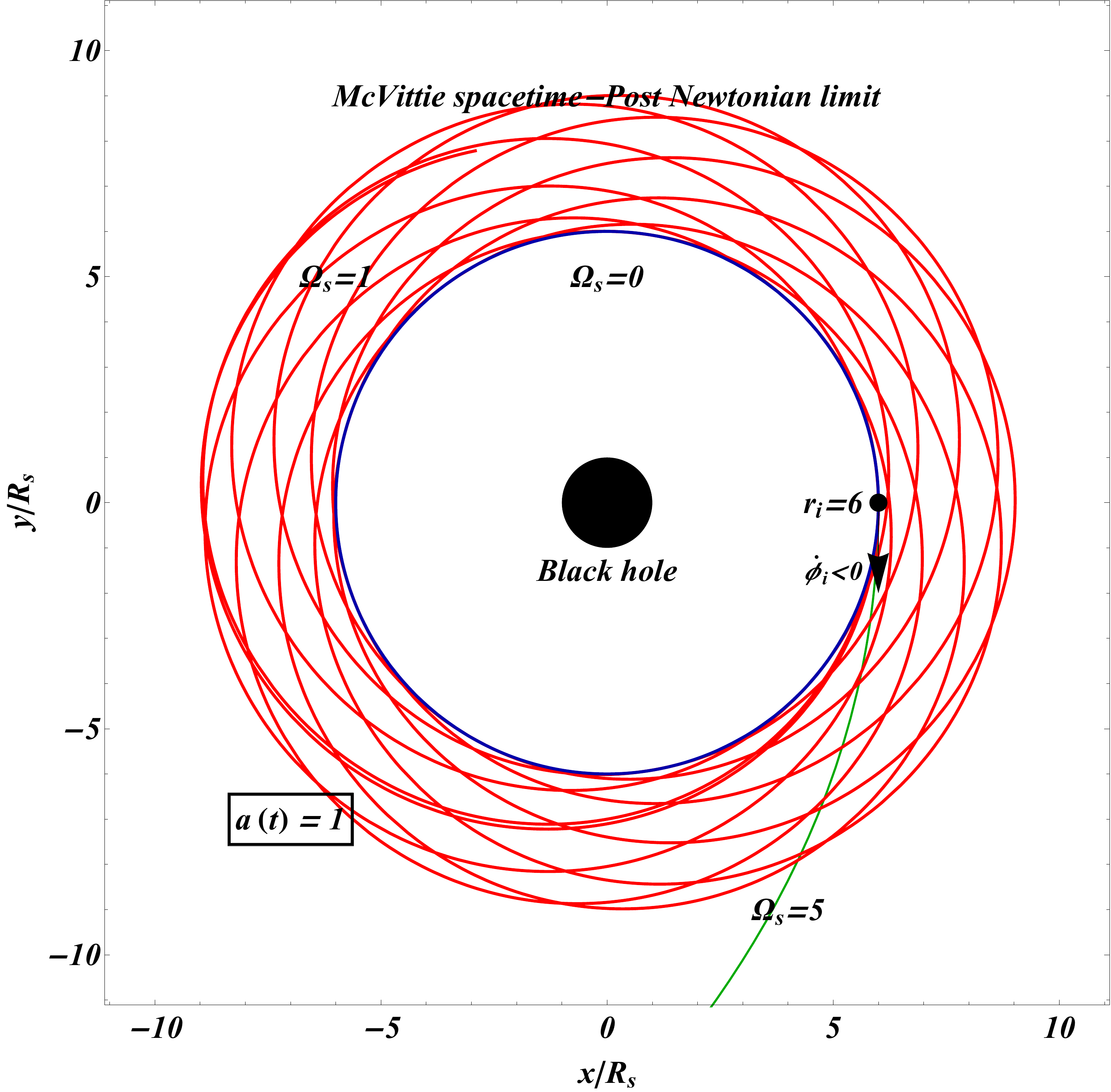} &
\epsfxsize=3.1in
\epsffile{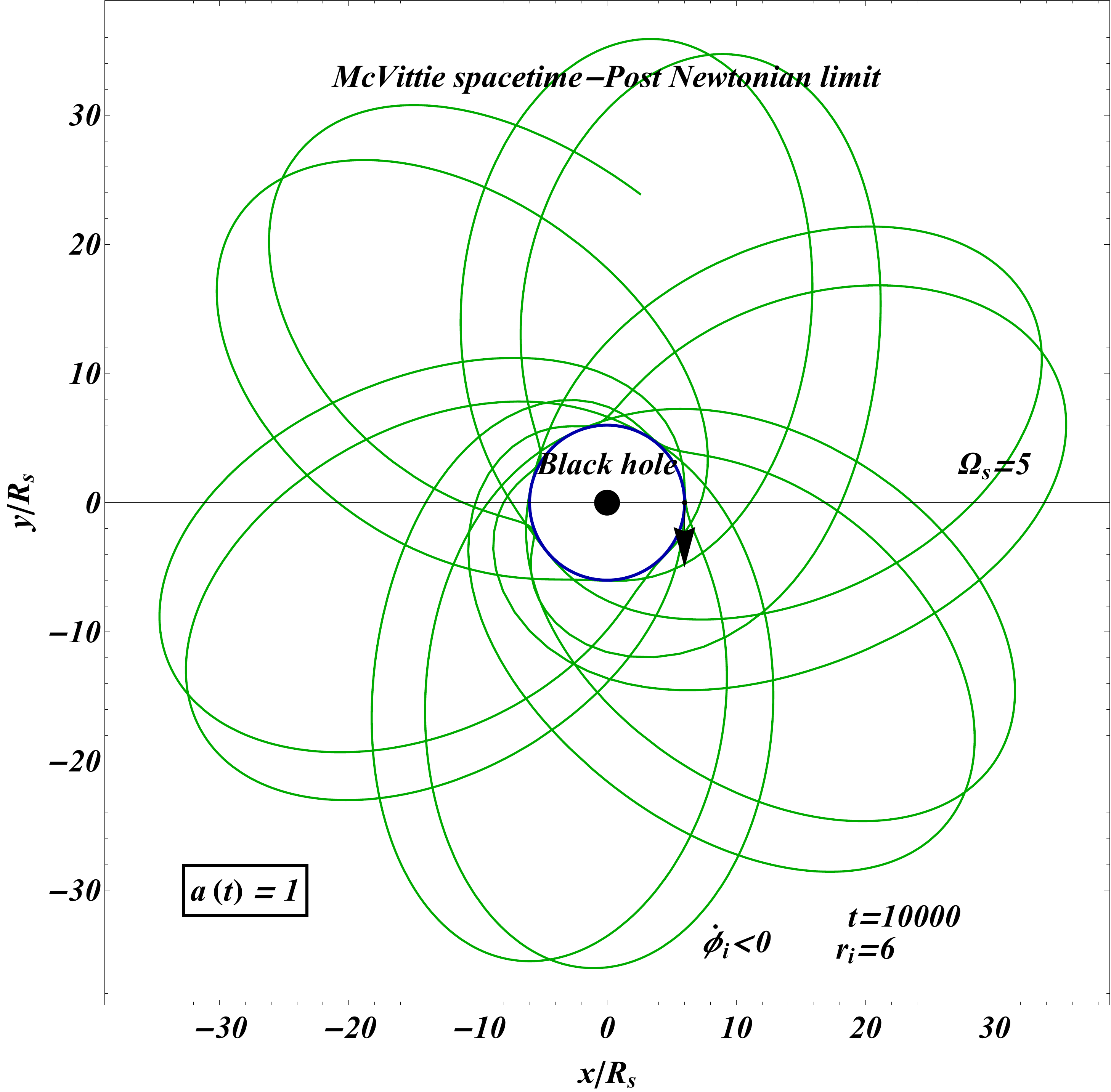} 
\end{array}$
\end{center}
\vspace{0.0cm}
\caption{Same as Fig. \ref{fig1} but the spinning particle orbits in the opposite direction. The circular orbits that would be present for a non-spinning particle get disrupted due to the spin-orbit coupling in the presence of spin. For $\Omega_s\dot{\phi}<0$ the spin-orbit coupling force is repulsive and the circular orbits are deformed outward. For non-spinning particle ($\Omega_s=0$) the circular orbits shown in both panels remain undisrupted. Notice that the $\Omega_s=0$ circular orbit, which corresponds to the absence of spin, is an inner bound for clockwise rotation. In any case the particle remains bound.}
\label{fig2}
\end{figure*}

Now, we introduce the rescalling through the variables $\bar{t}\equiv \frac{t}{R_s}$, $\bar{r}\equiv\frac{r}{R_s}$ and $\Omega_s \equiv \frac{\Omega}{R_s}=\frac{s}{\mu R_s}$ and from now on we omit the bar. The radial equation (\ref{mvr}) leads to \begin{equation}
\begin{aligned}
&\ddot{r}-\frac{\ddot{a}}{a}r-r\dot{\phi}^2=-r\Omega_s\dot\phi\bigg(\frac{\ddot a}{a}-\frac{\dot a^2}{a^2}+\frac{3}{2r^3}\bigg)+\\ &+\frac{1}{2}\bigg(-\frac{1}{r^2}-\dot\phi^2-\frac{\dot a^2}{a^2}+\frac{\dot r^2}{r^2}\bigg) \label{mvra}
\end{aligned}
\end{equation}

In the same way, the azimuthal equation (\ref{mvphi}) leads to \begin{equation}
\frac{d\big(r^2\dot{\phi}\big)}{dt}=\dot r\dot\phi+\Omega_s r^2\bigg(\frac{\dot r}{r}-\frac{\dot a}{a}\bigg)\bigg(\frac{\ddot a}{a}-\frac{\dot a^2}{a^2}\bigg)\label{mvphia}
\end{equation}

Equations (\ref{mvra}) and (\ref{mvphia}) are the main results of the present analysis. They generalize the geodesic equation of non-spinning particles in the post-Newtonian limit of McVittie metric and they reduce to those equations for $\Omega_s=0$. It is straightforward to solve numerically the system (\ref{mvra})-(\ref{mvphia}) and we implement such solutions in what follows. The following comments can be made on equations (\ref{mvra})-(\ref{mvphia}):
\begin{itemize}
\item
It is clear from Eq. (\ref{mvphia}) that the orbital angular momentum is not conserved due to the presence of the spin angular momentum. What is actually conserved is the $z$ component of the total angular momentum $J_z$ which is expressed through Eq. (\ref{f1a}) in terms of the angular and the spin angular momenta.
\item
The driving force term proportional to $\Omega_s$ and ${\dot \phi}$ in the radial geodesic equation (\ref{mvra}) has the form of a spin-orbit coupling and changes sign when the spin angular momentum reverses its direction with respect to the orbital angular momentum which is proportional to ${\dot\phi}$. This term is responsible for the deformation of the circular orbits and induces the well known chaotic behavior \cite{Verhaaren:2009md} of the spinning particle orbits in the absence of background expansion.
\end{itemize}

In what follows we solve the geodesic equations (\ref{mvra})-(\ref{mvphia}) for different forms of the expansion (static, accelerating, decelerating and constant) of the cosmological background and various values of the magnitude of the spin $s$ and consequently of the dimensionless parameter $\Omega_s$. We set  $\dot{r}(t_i)=0$  ($t_i=1$ is the initial time of the simulation) and $\dot\phi (t_i)$ so that $\ddot{r}_i=0$ corresponding to an initially circular orbit. We present analytically this issue in the Appendix. Also, we normalize the scale factor setting $a(1)=1$ and we set the particle at initial distance $r_i=6$ from the black hole.

\begin{subsection}{Numerical Solutions}

For a static universe ($a(t)=1$)  Eq. (\ref{mvra}) reduces to 

\begin{equation}
\ddot{r}=r\dot{\phi}^2+\frac{1}{2}(-\frac{1}{r^2}-\dot\phi^2+\frac{\dot r^2}{r^2}-\frac{3\Omega_s\dot\phi}{r^2}) \label{mvr1}
\end{equation}
while the Eq. (\ref{mvphia}) becomes
\begin{equation}
\frac{d(r^2\dot{\phi})}{dt}= \dot{r}\dot\phi\Rightarrow r^2\ddot{\phi}+(2r-1)\dot{r}\dot{\phi}=0 \label{mvphinew}
\end{equation}

The effect of the spin-orbit coupling force is demonstrated in Figs. \ref{fig1} and \ref{fig2} where we show the circular orbits disrupted due to the spin-orbit coupling. For $\Omega_s\dot{\phi}>0$ (see Fig. \ref{fig1})  the spin-orbit coupling force is attractive, since the term $-\frac{3\Omega_s\dot\phi}{2r^2}$ in Eq. (\ref{mvra}) is negative and the circular orbits (for a spinless particle) are deformed inward. The orbit of the motion of the particle remains bounded if the radius of the orbit is larger than $3R_s$. This is the well known effect of the 'innermost stable circular orbit' (ISCO) \cite{Barack:2009ey,Jefremov:2015gza,Suzuki:1997by}. It is defined as the smallest circular orbit in which a test particle can stably orbit a massive object \cite{Abramowicz:2010nk}. Since $r_{ISCO}=3R_s$ for a spinless central body in Schwarzschild spacetime, it is obvious that only black holes have innermost radius outside their surface.

This minimum allowed radius for bounded motion corresponds to a critical value of the dimensionless parameter $\Omega_s=0.6$ (left panel). Generally, in the presence of spin the orbits are bounded between a minimum and a maximum radius ($R_s=6$). As the spin increases ($\Omega_s>0.6$), at some time the orbit's radius becomes less than $3R_s$ and the particle gets captured by the black hole (right panel). For non-spinning particle ($\Omega_s=0$) the circular orbits shown in right panel remain undisrupted. 

\begin{figure*}[ht]
\centering
\begin{center}
$\begin{array}{@{\hspace{0.0in}}c@{\hspace{0.0in}}c@{\hspace{0.0in}}c}
\multicolumn{1}{l}{\mbox{}} 
\epsfxsize=3.4in
\epsffile{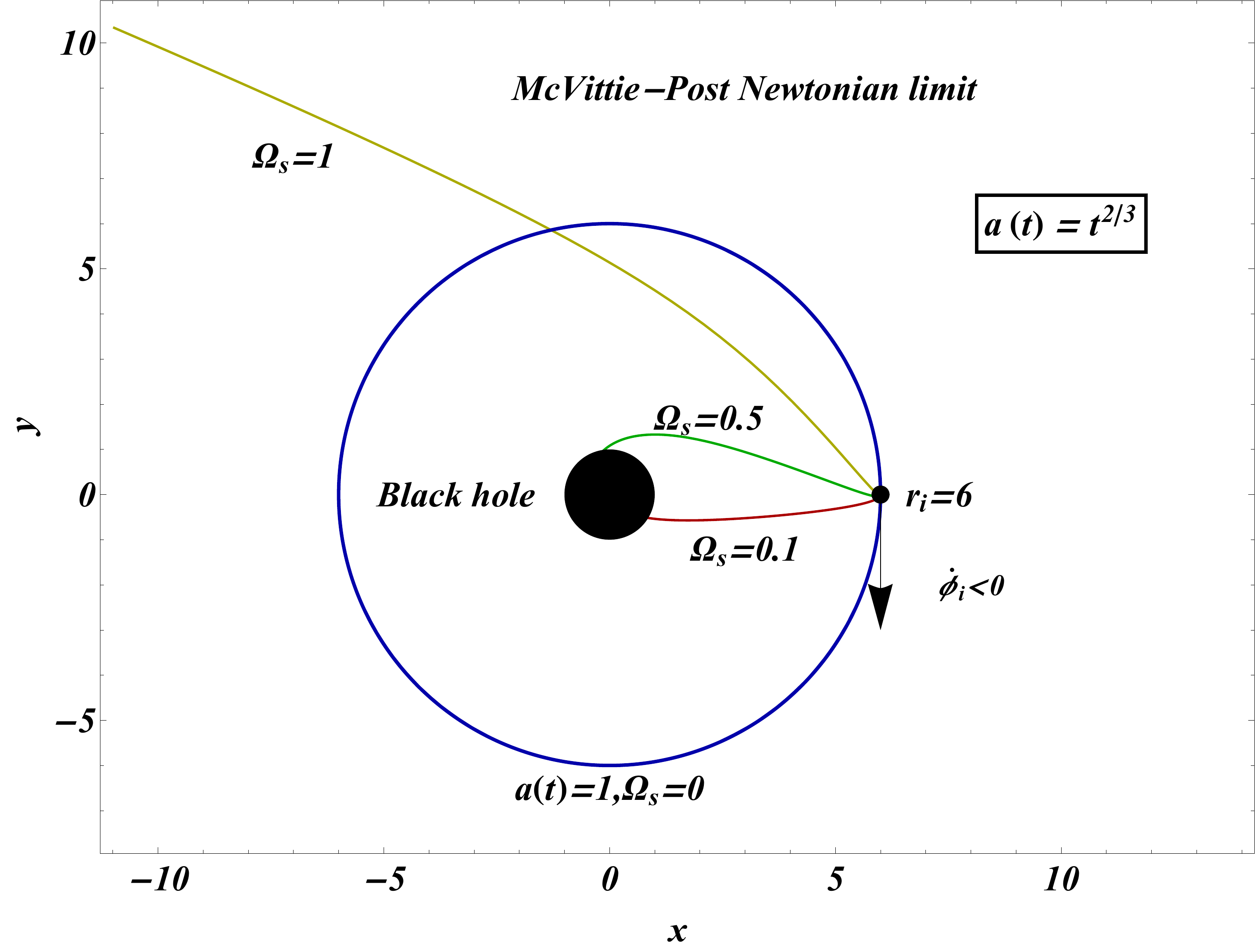} &
\epsfxsize=3in
\epsffile{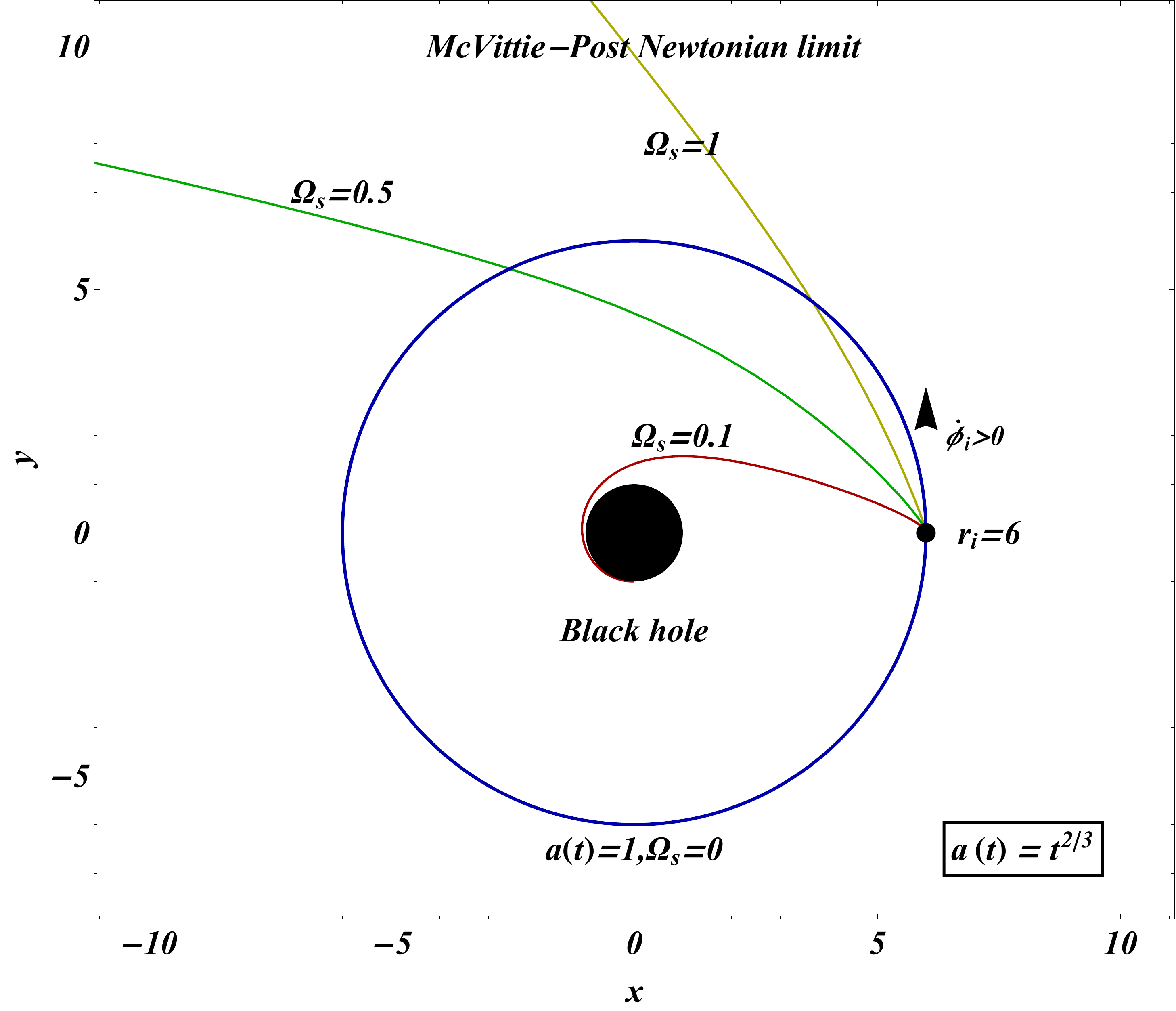} 
\end{array}$
\end{center}
\vspace{0.0cm}
\caption{The spinning particle orbits in the presence of decelerating universe expansion $a(t)\sim t^{2/3}$ for several values of the spin and initial clockwise (left panel with $\dot\phi(1)<0$) and counterclockwise (right panel with $\dot\phi(1)>0$) rotation. For small values of $\Omega_s$ the particle gets captured by the black hole, but as the parameter $\Omega_s$ increases the particle rapidly gets deflected to an unbounded orbit.}
\label{fig3}
\end{figure*} 

For $\Omega_s\dot{\phi}<0$ (see Fig. \ref{fig2})  the spin-orbit coupling force is repulsive, since the term $-\frac{3\Omega_s\dot\phi}{2r^2}$ in Eq. (\ref{mvra}) is positive and the circular orbits (for $s=0$) are deformed outward. The orbits of the motion of the spinning particle in all cases are bounded between a minimum ($R_s=6$) and a maximum radius.

In the presence of a decelerating expansion with $a(t)\sim t^{2/3}$ the orbits (solutions of Eqs. (\ref{mvra})-(\ref{mvphia})) are shown in Fig. \ref{fig3} for clockwise and counterclockwise rotation and initial conditions that would lead to a circular orbit in the absence of spin and expansion. In this case the effects of the expansion combined with the effects of the spin lead to rapid dissociation of the system or capture by the black hole. The result depends on the magnitude of the attractive and repulsive terms in Eq. (\ref{mvra}). Some orbits of the spinning particles for this case are shown in Fig. \ref{fig3}. 

In left panel of Fig. \ref{fig3} the initial rotation is clockwise, since $\dot\phi(1)<0$. In this case, the term $-\frac{3\Omega_s\dot\phi}{2r^2}$ in Eq. (\ref{mvra}) is repulsive and even if the cosmological background is decelerating, for large enough values of spin, such as $\Omega_s=1$ the particle rapidly gets deflected to an unbounded orbit. However, for small values of spin, such as $\Omega_s=0.1$ or $\Omega_s=0.5$ the decelerating background dominates and at some time the particle gets captured by the black hole. 

Similar results are shown in the right panel of Fig. \ref{fig3}, where the initial rotation of the particle is counterclockwise. In this case, the term $-\frac{3\Omega_s\dot\phi}{2r^2}$ in Eq. (\ref{mvra}) which describes the spin-orbit interaction is attractive. For small values of the dimensionless parameter $\Omega_s$, such as $\Omega_s=0.1$ the spinning particle approaches the black hole and when the radius of the orbit becomes less than $3R_s$, the particle gets captured by the strong gravity of the central body. However, when the spin takes larger values  such as $\Omega_s=0.5$ or $\Omega_s=1$ the particle gets deflected to an unbounded orbit, despite of the initially attractive effective force induced on the spinning particle. The expansion effects lead to dissociation of the initially bound system.

Now, we consider the effects of a de Sitter background expansion of the form \begin{equation}
a(t)=e^{Ht}\label{des}
\end{equation}
where $H=\sqrt{\frac{\bar\Lambda}{3}}$ and $\bar\Lambda$ is the cosmological constant in dimensionless form . We solve the system of Eqs. (\ref{mvra}) and (\ref{mvphia}) with the same initial conditions (circular orbit in the absence of spin and expansion). We set the cosmological constant equal to $\bar\Lambda=\Lambda R_s^2=3\times 10^{-2} $ \cite{Giblin:2003xj} and we present the trajectories of the particle in Fig. \ref{fig5}. We also show the corresponding orbit of a spinless particle in a static universe, in order to observe the deviation of each orbit from the circular. 

Setting a mass value of a typical black hole as $M=10M_\odot=2\times 10^{31}Kg$, we conclude that the dimensionless value $\Lambda R_s^2=0.03$ corresponds to  $\Lambda\simeq 3\times 10^6 sec^{-2}$ or $\Lambda\simeq 1.3\times 10^{-42}GeV^2$ much larger than the cosmological constant leading to the cosmic acceleration $\Lambda\simeq10^{-82}GeV^2$. Due to this normalization, orbit disturbances are much larger than the realistic form  corresponding to a  realistic cosmological setup. 

In left panel of Fig. \ref{fig5} the initial rotation is clockwise, since $\dot\phi(1)<0$. In this case, the term ($-\frac{3\Omega_s\dot\phi}{2r^2}$) in Eq. (\ref{mvra}) is positive and induces repulsion. In this case the repulsive effects of the accelerating cosmic expansion are amplified by the effects of the spin. 

For initial counterclockwise rotation (right panel in Fig. \ref{fig5}) the term  $-\frac{3\Omega_s\dot\phi}{2r^2}$ in radial equation is negative and induces attraction. However, for spinless particle or small values of spin and consequently of the parameter $\Omega_s$, such as $\Omega_s=10$, the accelerating cosmological background dominates and the particles get deflected to unbounded orbit. On the contrary, when the spin of the particle is large, such as $\Omega_s=100$, the attractive term $-\frac{3\Omega_s\dot\phi}{2r^2}$ in radial equation dominates the expansion and the spinning particle gets captured by the black hole.

\begin{figure*}[ht]
\centering
\begin{center}
$\begin{array}{@{\hspace{0.0in}}c@{\hspace{0.0in}}c@{\hspace{0.0in}}c}
\multicolumn{1}{l}{\mbox{}} 
\epsfxsize=2.8in
\epsffile{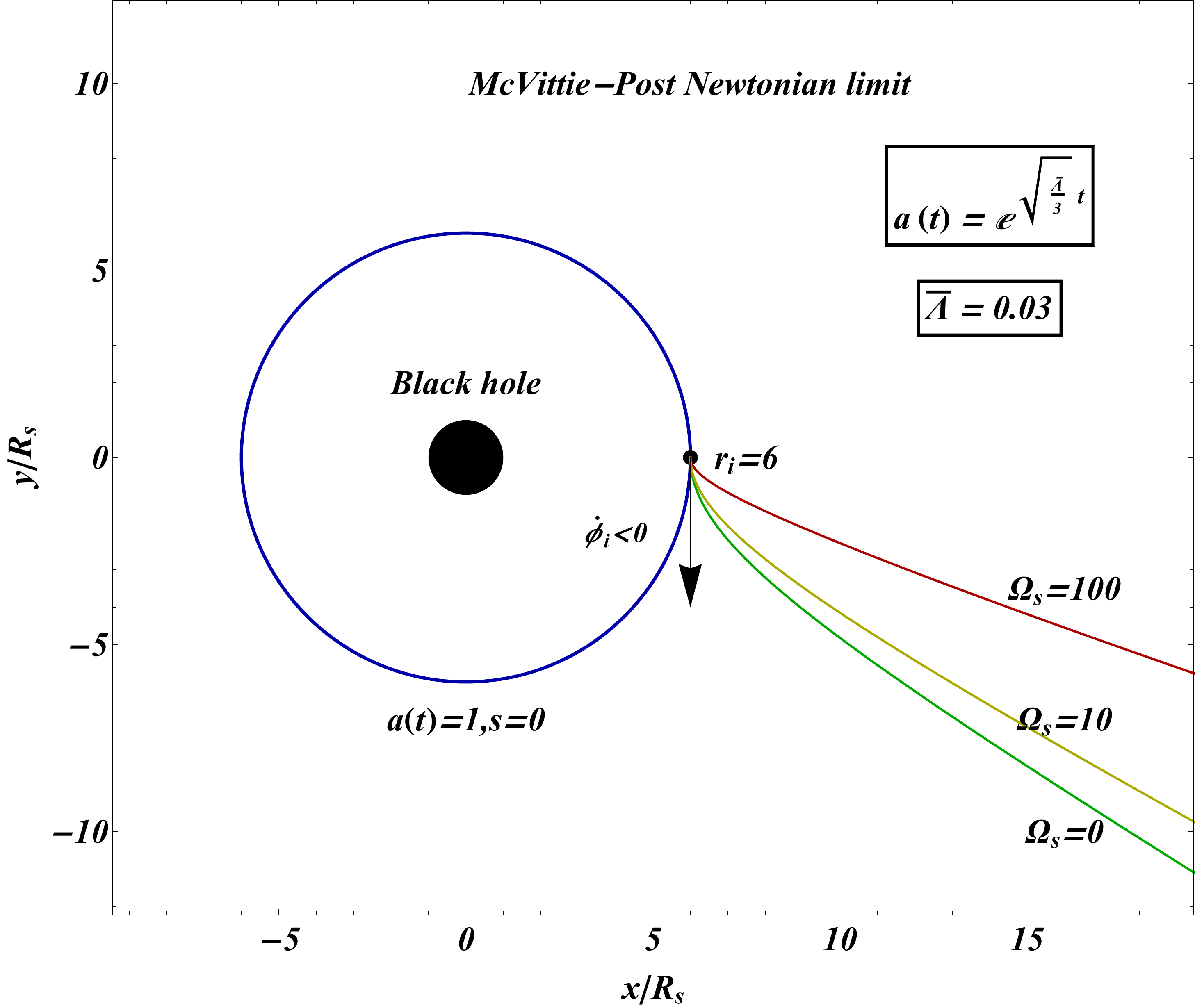} &
\epsfxsize=3.14in
\epsffile{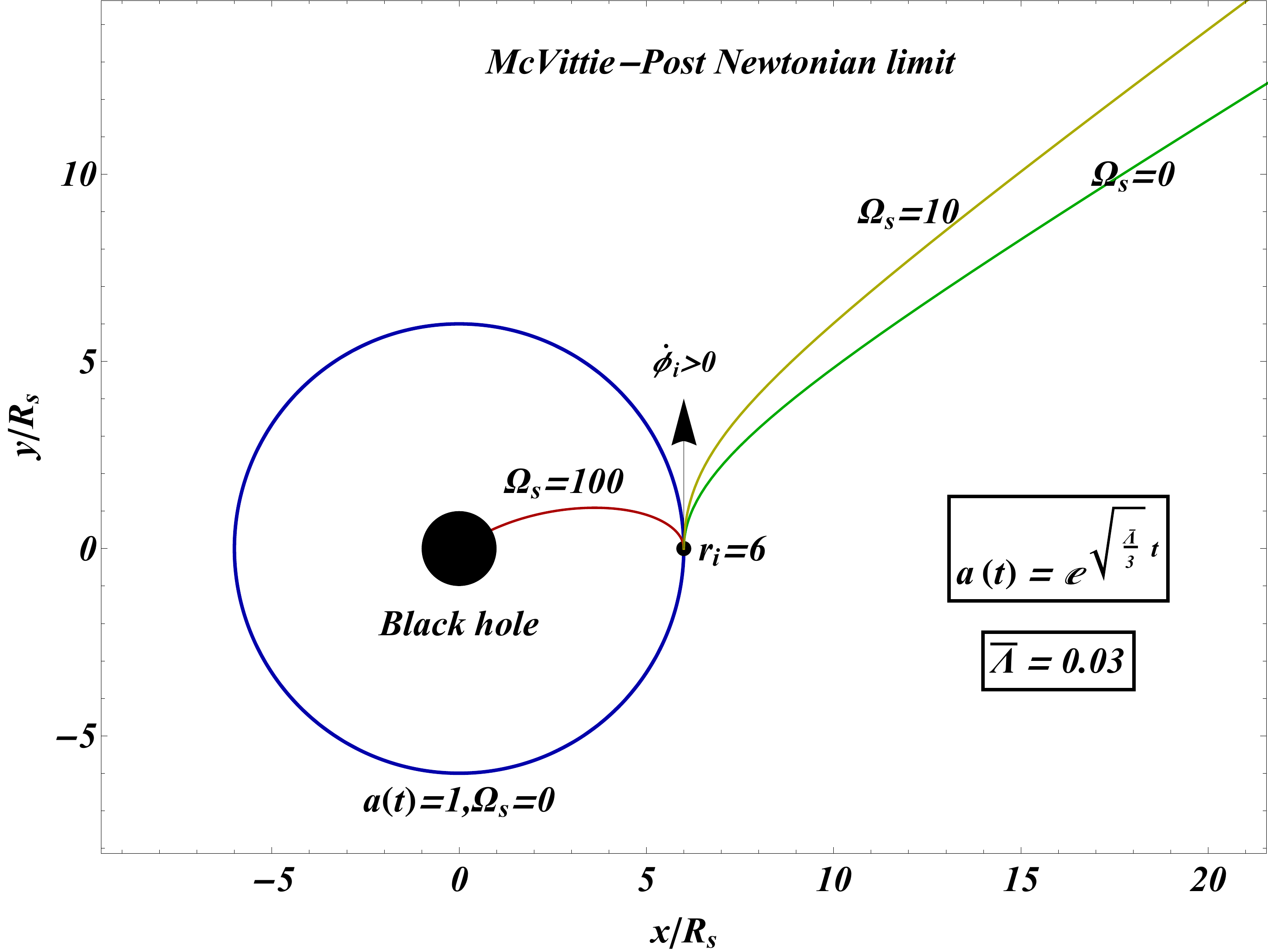} 
\end{array}$
\end{center}
\vspace{0.0cm}
\caption{ Same as Fig. \ref{fig3}, but the scale factor is of the form $a(t)=e^{\sqrt{\frac{\bar\Lambda}{3}}t}$ (de Sitter universe) with $\bar\Lambda=\Lambda R_s^2$. Notice the strong repulsive effects on the trajectories of the spinning/spinless particle for initial clockwise rotation (left panel) due to accelerating background expansion. The term $-\frac{3\Omega_s\dot\phi}{2r^2}$ in radial equation (\ref{mvra}) induces repulsion (left panel). However, for initial counterclockwise rotation (right panel) and extremely large spin, the particle captured by the black hole since the term $-\frac{3\Omega_s\dot\phi}{2r^2}$ in radial equation induces attraction and dominates.}
\label{fig5}
\end{figure*}

A crucial question of our analysis is which are the cosmological time intervals after which the effects of the expansion would become apparent. The answer can be easily obtained on dimensional grounds by equating the dimensionless parameters relevant for gravitational attraction ($M/r$) and background expansion $H_0\Delta t$ where $H_0$ is the Hubble parameter $\dot{a}/a$ at the present time and $\Delta t$ is the required time interval for the expansion effects to be observable. By equating these two parameters we find that the required time interval after which the cosmological expansion effects would become apparent on the trajectories is \begin{equation}
    \Delta t \simeq \frac{M}{H_0 r}\label{dt}
\end{equation} 
where we have set $G=1$. The time interval $\Delta t$ can be easily derived in S.I. as $\Delta t \simeq \frac{GM}{H_0 r c^2}$ and in Table \ref{table1} we give some estimates of the cosmological time intervals for a typical black hole, the solar system, a typical galaxy and a typical cluster of galaxies. The time intervals are in years, since we have consider that $1/H_0\simeq 1.4\times10^{10}$ years (the approximate age of the Universe).
\begin{table}[H]
\centering \scalebox{0.85}{
\begin{tabular}{ |c|c|c|c| } 
 \hline
  structure  & distance r ($m$) & mass M ($Kg$) & $\Delta t$ (years)\\
  \hline
 solar system & $5\times 10^{12}$  & $ 2\times 10^{30}$ & $\sim 4\times 10^0$ \\ 
 typical galaxy & $9\times10^{20}$ & $2\times10^{41}$ & $\sim 3\times 10^3$ \\ 
  cluster of galaxies & $3\times10^{22}$ & $2\times10^{45}$ & $\sim 7\times10^7$  \\
  black hole & $2\times10^{5}$ &$2\times10^{31}$ & $\sim 1\times10^9$  \\
 \hline
\end{tabular}}\caption{In this table we present estimations for some cosmological structures for the required time interval $\Delta t \simeq \frac{1}{H_0}\frac{GM}{rc^2}$ after which the cosmological expansion effects would become apparent on the trajectories. For the Hubble rate we have set $H_0^{-1}\simeq 1.4\times10^{10}$years. In the case of black hole we have consider the distance $r=6R_s$, as in the present work.}
\label{table1}
\end{table}

The MP equations have also been generalized to the case of modified theories of gravity, in which the matter energy-momentum tensor is not conserved. In modified gravity theories the Schwarzschild metric gets modified and so does the weak field limit, as we can see e.g. from eq. (32) of Ref. \cite{delaCruzDombriz:2009et}, which state to $f(R)$ theories ($G=1$)

\begin{equation}
    \chi(r)=1-\frac{2M}{r}+\frac{(1+f'(R_0))Q^2}{r^2}-\frac{R_0}{12}r^2\label{fr}
\end{equation}

Here, $Q=rV(r)$ is the charge of a black hole, $V(r)$ the potential and $R_0$ the curvature of the spacetime which we consider constant. An analysis along the line of the derivation of the McVittie metric for General Relativity (as discussed in \cite{Nolan:2014maa}) could generalize this metric to the case of $f(R)$ theories and also lead to the derivation of its Newtonian limit (the generalization of Eq. (\ref{mcvittief})). Alternatively one could directly include the scale factor $a(t)$ as a new factor along with the radial coordinate in Eq. (32) of \cite{delaCruzDombriz:2009et} and then take the Newtonian limit showing that it is a good approximation of the dynamical field equations for $f(R)$ gravity. This task is beyond the scope of the present analysis but it should be straightforward to implement in a future extension of our analysis.
\end{subsection}

\end{section}

\begin{section}{Conclusions} 

We have constructed and solved numerically the MP equations in the post Newtonian limit of McVittie background thus obtaining the orbits of spinning particles close to a massive object in an expanding cosmological background. We have identified the effects of a spin-orbit coupling which can be repulsive or attractive depending on the relative orientation between spin and orbital angular momentum. A static universe (no expansion) was shown to lead to disrupted spinning particle orbits which are not closed and are confined between a maximum and a minimum radius. This range increases with the value of the spin. As expected for the spin values, for which the radius of the motion of the particle becomes less that $3R_s$, the particle is captured by the black hole. This result is in agreement with previous studies that have indicated the presence of such behavior of the orbits \cite{Apostolatos}. 

Interesting extensions of our analysis include the construction and solution of the MP equations for the strong field regime of the McVittie metric, or the consideration of different SSC like the P condition.
\end{section}

\section*{Appendix}\label{A1}

In the present analysis we have focused on the distortion of orbits that would be circular in the absence of expansion and spin. In order to solve the system of equations (\ref{mvra}) and (\ref{mvphia}) we have assumed that initially the test particle has zero radial velocity ($\dot{r}(t_i=1)=0$)  and zero radial acceleration ($\ddot{r}(t_i=1)=0$). The initial value of the derivative $\dot{\phi}(1)$ can derived through the geodesic equation (\ref{mvra}). We set $a(t_i=1)=1$ and initial position for the particle $r_i=6$ in units of $R_s$.  Assuming a static Universe with $a(t)=1$ we compute the initial angular momentum from equation (\ref{mvra}). We set all the time derivatives of the scale factor equal to zero and thus we arrive at the following quadratic equation \be r_i^2(2r_i-1) (\dot{\phi}(1))^2-3\Omega_s\dot{\phi}(1)-1=0\label{mvphi1}\ee 
Setting $\Omega_s=0$, we obtain
\begin{equation}
    \dot{\phi}(1)=\pm\frac{\sqrt{11}}{66}\simeq \pm5\times 10^{-2}\label{roots}
\end{equation}

\raggedleft
\bibliography{spin}

\begin{thebibliography}{83}%
\makeatletter
\providecommand \@ifxundefined [1]{%
 \@ifx{#1\undefined}
}%
\providecommand \@ifnum [1]{%
 \ifnum #1\expandafter \@firstoftwo
 \else \expandafter \@secondoftwo
 \fi
}%
\providecommand \@ifx [1]{%
 \ifx #1\expandafter \@firstoftwo
 \else \expandafter \@secondoftwo
 \fi
}%
\providecommand \natexlab [1]{#1}%
\providecommand \enquote  [1]{``#1''}%
\providecommand \bibnamefont  [1]{#1}%
\providecommand \bibfnamefont [1]{#1}%
\providecommand \citenamefont [1]{#1}%
\providecommand \href@noop [0]{\@secondoftwo}%
\providecommand \href [0]{\begingroup \@sanitize@url \@href}%
\providecommand \@href[1]{\@@startlink{#1}\@@href}%
\providecommand \@@href[1]{\endgroup#1\@@endlink}%
\providecommand \@sanitize@url [0]{\catcode `\\12\catcode `\$12\catcode
  `\&12\catcode `\#12\catcode `\^12\catcode `\_12\catcode `\%12\relax}%
\providecommand \@@startlink[1]{}%
\providecommand \@@endlink[0]{}%
\providecommand \url  [0]{\begingroup\@sanitize@url \@url }%
\providecommand \@url [1]{\endgroup\@href {#1}{\urlprefix }}%
\providecommand \urlprefix  [0]{URL }%
\providecommand \Eprint [0]{\href }%
\providecommand \doibase [0]{http://dx.doi.org/}%
\providecommand \selectlanguage [0]{\@gobble}%
\providecommand \bibinfo  [0]{\@secondoftwo}%
\providecommand \bibfield  [0]{\@secondoftwo}%
\providecommand \translation [1]{[#1]}%
\providecommand \BibitemOpen [0]{}%
\providecommand \bibitemStop [0]{}%
\providecommand \bibitemNoStop [0]{.\EOS\space}%
\providecommand \EOS [0]{\spacefactor3000\relax}%
\providecommand \BibitemShut  [1]{\csname bibitem#1\endcsname}%
\let\auto@bib@innerbib\@empty
\bibitem [{\citenamefont {Hartl}(2003{\natexlab{a}})}]{Hartl:2002ig}%
  \BibitemOpen
  \bibfield  {author} {\bibinfo {author} {\bibfnamefont {Michael~D.}\
  \bibnamefont {Hartl}},\ }\bibfield  {title} {\enquote {\bibinfo {title}
  {{Dynamics of spinning test particles in Kerr space-time}},}\ }\href
  {\doibase 10.1103/PhysRevD.67.024005} {\bibfield  {journal} {\bibinfo
  {journal} {Phys. Rev.}\ }\textbf {\bibinfo {volume} {D67}},\ \bibinfo {pages}
  {024005} (\bibinfo {year} {2003}{\natexlab{a}})},\ \Eprint
  {http://arxiv.org/abs/gr-qc/0210042} {arXiv:gr-qc/0210042 [gr-qc]}
  \BibitemShut {NoStop}%
\bibitem [{\citenamefont {Caldwell}\ \emph {et~al.}(2003)\citenamefont
  {Caldwell}, \citenamefont {Kamionkowski},\ and\ \citenamefont
  {Weinberg}}]{Caldwell:2003vq}%
  \BibitemOpen
  \bibfield  {author} {\bibinfo {author} {\bibfnamefont {Robert~R.}\
  \bibnamefont {Caldwell}}, \bibinfo {author} {\bibfnamefont {Marc}\
  \bibnamefont {Kamionkowski}}, \ and\ \bibinfo {author} {\bibfnamefont
  {Nevin~N.}\ \bibnamefont {Weinberg}},\ }\bibfield  {title} {\enquote
  {\bibinfo {title} {{Phantom energy and cosmic doomsday}},}\ }\href {\doibase
  10.1103/PhysRevLett.91.071301} {\bibfield  {journal} {\bibinfo  {journal}
  {Phys. Rev. Lett.}\ }\textbf {\bibinfo {volume} {91}},\ \bibinfo {pages}
  {071301} (\bibinfo {year} {2003})},\ \Eprint
  {http://arxiv.org/abs/astro-ph/0302506} {arXiv:astro-ph/0302506 [astro-ph]}
  \BibitemShut {NoStop}%
\bibitem [{\citenamefont {Nesseris}\ and\ \citenamefont
  {Perivolaropoulos}(2004)}]{Nesseris:2004uj}%
  \BibitemOpen
  \bibfield  {author} {\bibinfo {author} {\bibfnamefont {S.}~\bibnamefont
  {Nesseris}}\ and\ \bibinfo {author} {\bibfnamefont {Leandros}\ \bibnamefont
  {Perivolaropoulos}},\ }\bibfield  {title} {\enquote {\bibinfo {title} {{The
  Fate of bound systems in phantom and quintessence cosmologies}},}\ }\href
  {\doibase 10.1103/PhysRevD.70.123529} {\bibfield  {journal} {\bibinfo
  {journal} {Phys. Rev.}\ }\textbf {\bibinfo {volume} {D70}},\ \bibinfo {pages}
  {123529} (\bibinfo {year} {2004})},\ \Eprint
  {http://arxiv.org/abs/astro-ph/0410309} {arXiv:astro-ph/0410309 [astro-ph]}
  \BibitemShut {NoStop}%
\bibitem [{\citenamefont {Antoniou}\ and\ \citenamefont
  {Perivolaropoulos}(2016)}]{Antoniou:2016obw}%
  \BibitemOpen
  \bibfield  {author} {\bibinfo {author} {\bibfnamefont {Ioannis}\ \bibnamefont
  {Antoniou}}\ and\ \bibinfo {author} {\bibfnamefont {Leandros}\ \bibnamefont
  {Perivolaropoulos}},\ }\bibfield  {title} {\enquote {\bibinfo {title}
  {{Geodesics of McVittie Spacetime with a Phantom Cosmological Background}},}\
  }\href {\doibase 10.1103/PhysRevD.93.123520} {\bibfield  {journal} {\bibinfo
  {journal} {Phys. Rev.}\ }\textbf {\bibinfo {volume} {D93}},\ \bibinfo {pages}
  {123520} (\bibinfo {year} {2016})},\ \Eprint
  {http://arxiv.org/abs/1603.02569} {arXiv:1603.02569 [gr-qc]} \BibitemShut
  {NoStop}%
\bibitem [{\citenamefont {Nolan}(2014)}]{Nolan:2014maa}%
  \BibitemOpen
  \bibfield  {author} {\bibinfo {author} {\bibfnamefont {Brien~C.}\
  \bibnamefont {Nolan}},\ }\bibfield  {title} {\enquote {\bibinfo {title}
  {{Particle and photon orbits in McVittie spacetimes}},}\ }\href {\doibase
  10.1088/0264-9381/31/23/235008} {\bibfield  {journal} {\bibinfo  {journal}
  {Class. Quant. Grav.}\ }\textbf {\bibinfo {volume} {31}},\ \bibinfo {pages}
  {235008} (\bibinfo {year} {2014})},\ \Eprint {http://arxiv.org/abs/1408.0044}
  {arXiv:1408.0044 [gr-qc]} \BibitemShut {NoStop}%
\bibitem [{\citenamefont {Obukhov}\ \emph {et~al.}(2013)\citenamefont
  {Obukhov}, \citenamefont {Silenko},\ and\ \citenamefont
  {Teryaev}}]{Obukhov:2013zca}%
  \BibitemOpen
  \bibfield  {author} {\bibinfo {author} {\bibfnamefont {Yuri~N.}\ \bibnamefont
  {Obukhov}}, \bibinfo {author} {\bibfnamefont {Alexander~J.}\ \bibnamefont
  {Silenko}}, \ and\ \bibinfo {author} {\bibfnamefont {Oleg~V.}\ \bibnamefont
  {Teryaev}},\ }\bibfield  {title} {\enquote {\bibinfo {title} {{Spin in an
  arbitrary gravitational field}},}\ }\href {\doibase
  10.1103/PhysRevD.88.084014} {\bibfield  {journal} {\bibinfo  {journal} {Phys.
  Rev.}\ }\textbf {\bibinfo {volume} {D88}},\ \bibinfo {pages} {084014}
  (\bibinfo {year} {2013})},\ \Eprint {http://arxiv.org/abs/1308.4552}
  {arXiv:1308.4552 [gr-qc]} \BibitemShut {NoStop}%
\bibitem [{\citenamefont {Tod}\ \emph {et~al.}(1976)\citenamefont {Tod},
  \citenamefont {de~Felice},\ and\ \citenamefont {Calvani}}]{Tod:1976ud}%
  \BibitemOpen
  \bibfield  {author} {\bibinfo {author} {\bibfnamefont {K.~P.}\ \bibnamefont
  {Tod}}, \bibinfo {author} {\bibfnamefont {F.}~\bibnamefont {de~Felice}}, \
  and\ \bibinfo {author} {\bibfnamefont {M.}~\bibnamefont {Calvani}},\
  }\bibfield  {title} {\enquote {\bibinfo {title} {{Spinning test particles in
  the field of a black hole}},}\ }\href {\doibase 10.1007/BF02728614}
  {\bibfield  {journal} {\bibinfo  {journal} {Nuovo Cim.}\ }\textbf {\bibinfo
  {volume} {B34}},\ \bibinfo {pages} {365} (\bibinfo {year}
  {1976})}\BibitemShut {NoStop}%
\bibitem [{\citenamefont {Semerak}(1999)}]{Semerak:1999qc}%
  \BibitemOpen
  \bibfield  {author} {\bibinfo {author} {\bibfnamefont {O.}~\bibnamefont
  {Semerak}},\ }\bibfield  {title} {\enquote {\bibinfo {title} {{Spinning test
  particles in a Kerr field. 1.}}}\ }\href {\doibase
  10.1046/j.1365-8711.1999.02754.x} {\bibfield  {journal} {\bibinfo  {journal}
  {Mon. Not. Roy. Astron. Soc.}\ }\textbf {\bibinfo {volume} {308}},\ \bibinfo
  {pages} {863--875} (\bibinfo {year} {1999})}\BibitemShut {NoStop}%
\bibitem [{\citenamefont {Khriplovich}(2008)}]{Khriplovich:2008ni}%
  \BibitemOpen
  \bibfield  {author} {\bibinfo {author} {\bibfnamefont {I.~B.}\ \bibnamefont
  {Khriplovich}},\ }\bibfield  {title} {\enquote {\bibinfo {title} {{Spinning
  Relativistic Particles in External Fields}},}\ }\href@noop {} {\  (\bibinfo
  {year} {2008})},\ \Eprint {http://arxiv.org/abs/0801.1881} {arXiv:0801.1881
  [gr-qc]} \BibitemShut {NoStop}%
\bibitem [{\citenamefont
  {Lukes-Gerakopoulos}(2017{\natexlab{a}})}]{Lukes-Gerakopoulos:2016udm}%
  \BibitemOpen
  \bibfield  {author} {\bibinfo {author} {\bibfnamefont {Georgios}\
  \bibnamefont {Lukes-Gerakopoulos}},\ }\bibfield  {title} {\enquote {\bibinfo
  {title} {{Spinning particles moving around black holes: integrability and
  chaos}},}\ }in\ \href {\doibase 10.1142/9789813226609_0209} {\emph {\bibinfo
  {booktitle} {{Proceedings, 14th Marcel Grossmann Meeting on Recent
  Developments in Theoretical and Experimental General Relativity,
  Astrophysics, and Relativistic Field Theories (MG14) (In 4 Volumes): Rome,
  Italy, July 12-18, 2015}}}},\ Vol.~\bibinfo {volume} {2}\ (\bibinfo {year}
  {2017})\ pp.\ \bibinfo {pages} {1960--1965},\ \Eprint
  {http://arxiv.org/abs/1606.09430} {arXiv:1606.09430 [gr-qc]} \BibitemShut
  {NoStop}%
\bibitem [{\citenamefont {Costa}\ \emph {et~al.}(2016)\citenamefont {Costa},
  \citenamefont {Natário},\ and\ \citenamefont {Zilhao}}]{Costa:2012cy}%
  \BibitemOpen
  \bibfield  {author} {\bibinfo {author} {\bibfnamefont {L.~Filipe~O.}\
  \bibnamefont {Costa}}, \bibinfo {author} {\bibfnamefont {José}\ \bibnamefont
  {Natário}}, \ and\ \bibinfo {author} {\bibfnamefont {Miguel}\ \bibnamefont
  {Zilhao}},\ }\bibfield  {title} {\enquote {\bibinfo {title} {{Spacetime
  dynamics of spinning particles: Exact electromagnetic analogies}},}\ }\href
  {\doibase 10.1103/PhysRevD.93.104006} {\bibfield  {journal} {\bibinfo
  {journal} {Phys. Rev.}\ }\textbf {\bibinfo {volume} {D93}},\ \bibinfo {pages}
  {104006} (\bibinfo {year} {2016})},\ \Eprint {http://arxiv.org/abs/1207.0470}
  {arXiv:1207.0470 [gr-qc]} \BibitemShut {NoStop}%
\bibitem [{\citenamefont {Burko}(2004)}]{Burko:2003rv}%
  \BibitemOpen
  \bibfield  {author} {\bibinfo {author} {\bibfnamefont {Lior~M.}\ \bibnamefont
  {Burko}},\ }\bibfield  {title} {\enquote {\bibinfo {title} {{Orbital
  evolution of a particle around a black hole. 2. Comparison of contributions
  of spin orbit coupling and the selfforce}},}\ }\href {\doibase
  10.1103/PhysRevD.69.044011} {\bibfield  {journal} {\bibinfo  {journal} {Phys.
  Rev.}\ }\textbf {\bibinfo {volume} {D69}},\ \bibinfo {pages} {044011}
  (\bibinfo {year} {2004})},\ \Eprint {http://arxiv.org/abs/gr-qc/0308003}
  {arXiv:gr-qc/0308003 [gr-qc]} \BibitemShut {NoStop}%
\bibitem [{\citenamefont {Shibata}(1993)}]{Shibata:1993uk}%
  \BibitemOpen
  \bibfield  {author} {\bibinfo {author} {\bibfnamefont {Masaru}\ \bibnamefont
  {Shibata}},\ }\bibfield  {title} {\enquote {\bibinfo {title} {{Gravitational
  waves induced by a particle orbiting around a rotating black hole: spin orbit
  interaction effect}},}\ }\href {\doibase 10.1103/PhysRevD.48.663} {\bibfield
  {journal} {\bibinfo  {journal} {Phys. Rev.}\ }\textbf {\bibinfo {volume}
  {D48}},\ \bibinfo {pages} {663--666} (\bibinfo {year} {1993})}\BibitemShut
  {NoStop}%
\bibitem [{\citenamefont {Han}\ and\ \citenamefont {Yang}(2017)}]{Han:2016djt}%
  \BibitemOpen
  \bibfield  {author} {\bibinfo {author} {\bibfnamefont {Wen-Bias}\
  \bibnamefont {Han}}\ and\ \bibinfo {author} {\bibfnamefont {Shu-Cheng}\
  \bibnamefont {Yang}},\ }\bibfield  {title} {\enquote {\bibinfo {title}
  {{Exotic orbits due to spin–spin coupling around Kerr black holes}},}\
  }\href {\doibase 10.1142/S0218271817501796} {\bibfield  {journal} {\bibinfo
  {journal} {Int. J. Mod. Phys.}\ }\textbf {\bibinfo {volume} {D27}},\ \bibinfo
  {pages} {1750179} (\bibinfo {year} {2017})},\ \Eprint
  {http://arxiv.org/abs/1610.01534} {arXiv:1610.01534 [gr-qc]} \BibitemShut
  {NoStop}%
\bibitem [{\citenamefont {Suzuki}\ and\ \citenamefont
  {Maeda}(2000)}]{Suzuki:1999si}%
  \BibitemOpen
  \bibfield  {author} {\bibinfo {author} {\bibfnamefont {Shingo}\ \bibnamefont
  {Suzuki}}\ and\ \bibinfo {author} {\bibfnamefont {Kei-ichi}\ \bibnamefont
  {Maeda}},\ }\bibfield  {title} {\enquote {\bibinfo {title} {{Signature of
  chaos in gravitational waves from a spinning particle}},}\ }\href {\doibase
  10.1103/PhysRevD.61.024005} {\bibfield  {journal} {\bibinfo  {journal} {Phys.
  Rev.}\ }\textbf {\bibinfo {volume} {D61}},\ \bibinfo {pages} {024005}
  (\bibinfo {year} {2000})},\ \Eprint {http://arxiv.org/abs/gr-qc/9910064}
  {arXiv:gr-qc/9910064 [gr-qc]} \BibitemShut {NoStop}%
\bibitem [{\citenamefont {Kubiznak}\ and\ \citenamefont
  {Cariglia}(2012)}]{Kubiznak:2011ay}%
  \BibitemOpen
  \bibfield  {author} {\bibinfo {author} {\bibfnamefont {David}\ \bibnamefont
  {Kubiznak}}\ and\ \bibinfo {author} {\bibfnamefont {Marco}\ \bibnamefont
  {Cariglia}},\ }\bibfield  {title} {\enquote {\bibinfo {title} {{On
  Integrability of spinning particle motion in higher-dimensional black hole
  spacetimes}},}\ }\href {\doibase 10.1103/PhysRevLett.108.051104} {\bibfield
  {journal} {\bibinfo  {journal} {Phys. Rev. Lett.}\ }\textbf {\bibinfo
  {volume} {108}},\ \bibinfo {pages} {051104} (\bibinfo {year} {2012})},\
  \Eprint {http://arxiv.org/abs/1110.0495} {arXiv:1110.0495 [hep-th]}
  \BibitemShut {NoStop}%
\bibitem [{\citenamefont {Kao}\ and\ \citenamefont {Cho}(2005)}]{Kao:2004qs}%
  \BibitemOpen
  \bibfield  {author} {\bibinfo {author} {\bibfnamefont {J.~K.}\ \bibnamefont
  {Kao}}\ and\ \bibinfo {author} {\bibfnamefont {H.~T.}\ \bibnamefont {Cho}},\
  }\bibfield  {title} {\enquote {\bibinfo {title} {{The Onset of chaotic motion
  of a spinning particle around the Schwarzchild black hole}},}\ }\href
  {\doibase 10.1016/j.physleta.2005.01.020} {\bibfield  {journal} {\bibinfo
  {journal} {Phys. Lett.}\ }\textbf {\bibinfo {volume} {A336}},\ \bibinfo
  {pages} {159--166} (\bibinfo {year} {2005})},\ \Eprint
  {http://arxiv.org/abs/gr-qc/0406101} {arXiv:gr-qc/0406101 [gr-qc]}
  \BibitemShut {NoStop}%
\bibitem [{\citenamefont {Saijo}\ \emph {et~al.}(1998)\citenamefont {Saijo},
  \citenamefont {Maeda}, \citenamefont {Shibata},\ and\ \citenamefont
  {Mino}}]{Saijo:1998mn}%
  \BibitemOpen
  \bibfield  {author} {\bibinfo {author} {\bibfnamefont {Motoyuki}\
  \bibnamefont {Saijo}}, \bibinfo {author} {\bibfnamefont {Kei-ichi}\
  \bibnamefont {Maeda}}, \bibinfo {author} {\bibfnamefont {Masaru}\
  \bibnamefont {Shibata}}, \ and\ \bibinfo {author} {\bibfnamefont {Yashushi}\
  \bibnamefont {Mino}},\ }\bibfield  {title} {\enquote {\bibinfo {title}
  {{Gravitational waves from a spinning particle plunging into a Kerr black
  hole}},}\ }\href {\doibase 10.1103/PhysRevD.58.064005} {\bibfield  {journal}
  {\bibinfo  {journal} {Phys. Rev.}\ }\textbf {\bibinfo {volume} {D58}},\
  \bibinfo {pages} {064005} (\bibinfo {year} {1998})}\BibitemShut {NoStop}%
\bibitem [{\citenamefont {Tanaka}\ \emph {et~al.}(1996)\citenamefont {Tanaka},
  \citenamefont {Mino}, \citenamefont {Sasaki},\ and\ \citenamefont
  {Shibata}}]{Tanaka:1996ht}%
  \BibitemOpen
  \bibfield  {author} {\bibinfo {author} {\bibfnamefont {Takahiro}\
  \bibnamefont {Tanaka}}, \bibinfo {author} {\bibfnamefont {Yasushi}\
  \bibnamefont {Mino}}, \bibinfo {author} {\bibfnamefont {Misao}\ \bibnamefont
  {Sasaki}}, \ and\ \bibinfo {author} {\bibfnamefont {Masaru}\ \bibnamefont
  {Shibata}},\ }\bibfield  {title} {\enquote {\bibinfo {title} {{Gravitational
  waves from a spinning particle in circular orbits around a rotating black
  hole}},}\ }\href {\doibase 10.1103/PhysRevD.54.3762} {\bibfield  {journal}
  {\bibinfo  {journal} {Phys. Rev.}\ }\textbf {\bibinfo {volume} {D54}},\
  \bibinfo {pages} {3762--3777} (\bibinfo {year} {1996})},\ \Eprint
  {http://arxiv.org/abs/gr-qc/9602038} {arXiv:gr-qc/9602038 [gr-qc]}
  \BibitemShut {NoStop}%
\bibitem [{\citenamefont {Mino}\ \emph {et~al.}(1996)\citenamefont {Mino},
  \citenamefont {Shibata},\ and\ \citenamefont {Tanaka}}]{Mino:1995fm}%
  \BibitemOpen
  \bibfield  {author} {\bibinfo {author} {\bibfnamefont {Yasushi}\ \bibnamefont
  {Mino}}, \bibinfo {author} {\bibfnamefont {Masaru}\ \bibnamefont {Shibata}},
  \ and\ \bibinfo {author} {\bibfnamefont {Takahiro}\ \bibnamefont {Tanaka}},\
  }\bibfield  {title} {\enquote {\bibinfo {title} {{Gravitational waves induced
  by a spinning particle falling into a rotating black hole}},}\ }\href
  {\doibase 10.1103/PhysRevD.53.622, 10.1103/PhysRevD.59.047502} {\bibfield
  {journal} {\bibinfo  {journal} {Phys. Rev.}\ }\textbf {\bibinfo {volume}
  {D53}},\ \bibinfo {pages} {622--634} (\bibinfo {year} {1996})},\ \bibinfo
  {note} {[Erratum: Phys. Rev.D59,047502(1999)]}\BibitemShut {NoStop}%
\bibitem [{\citenamefont {Harms}\ \emph {et~al.}(2016)\citenamefont {Harms},
  \citenamefont {Lukes-Gerakopoulos}, \citenamefont {Bernuzzi},\ and\
  \citenamefont {Nagar}}]{Harms:2016ctx}%
  \BibitemOpen
  \bibfield  {author} {\bibinfo {author} {\bibfnamefont {Enno}\ \bibnamefont
  {Harms}}, \bibinfo {author} {\bibfnamefont {Georgios}\ \bibnamefont
  {Lukes-Gerakopoulos}}, \bibinfo {author} {\bibfnamefont {Sebastiano}\
  \bibnamefont {Bernuzzi}}, \ and\ \bibinfo {author} {\bibfnamefont
  {Alessandro}\ \bibnamefont {Nagar}},\ }\bibfield  {title} {\enquote {\bibinfo
  {title} {{Spinning test body orbiting around a Schwarzschild black hole:
  Circular dynamics and gravitational-wave fluxes}},}\ }\href {\doibase
  10.1103/PhysRevD.94.104010} {\bibfield  {journal} {\bibinfo  {journal} {Phys.
  Rev.}\ }\textbf {\bibinfo {volume} {D94}},\ \bibinfo {pages} {104010}
  (\bibinfo {year} {2016})},\ \Eprint {http://arxiv.org/abs/1609.00356}
  {arXiv:1609.00356 [gr-qc]} \BibitemShut {NoStop}%
\bibitem [{\citenamefont {Kaloper}\ \emph {et~al.}(2010)\citenamefont
  {Kaloper}, \citenamefont {Kleban},\ and\ \citenamefont
  {Martin}}]{Kaloper:2010ec}%
  \BibitemOpen
  \bibfield  {author} {\bibinfo {author} {\bibfnamefont {Nemanja}\ \bibnamefont
  {Kaloper}}, \bibinfo {author} {\bibfnamefont {Matthew}\ \bibnamefont
  {Kleban}}, \ and\ \bibinfo {author} {\bibfnamefont {Damien}\ \bibnamefont
  {Martin}},\ }\bibfield  {title} {\enquote {\bibinfo {title} {{McVittie's
  Legacy: Black Holes in an Expanding Universe}},}\ }\href {\doibase
  10.1103/PhysRevD.81.104044} {\bibfield  {journal} {\bibinfo  {journal} {Phys.
  Rev.}\ }\textbf {\bibinfo {volume} {D81}},\ \bibinfo {pages} {104044}
  (\bibinfo {year} {2010})},\ \Eprint {http://arxiv.org/abs/1003.4777}
  {arXiv:1003.4777 [hep-th]} \BibitemShut {NoStop}%
\bibitem [{\citenamefont {Mathisson}(1937)}]{Mathisson:1937zz}%
  \BibitemOpen
  \bibfield  {author} {\bibinfo {author} {\bibfnamefont {Myron}\ \bibnamefont
  {Mathisson}},\ }\bibfield  {title} {\enquote {\bibinfo {title} {{Neue
  mechanik materieller systemes}},}\ }\href@noop {} {\bibfield  {journal}
  {\bibinfo  {journal} {Acta Phys. Polon.}\ }\textbf {\bibinfo {volume} {6}},\
  \bibinfo {pages} {163--2900} (\bibinfo {year} {1937})}\BibitemShut {NoStop}%
\bibitem [{\citenamefont {Papapetrou}(1951)}]{Papapetrou:1951pa}%
  \BibitemOpen
  \bibfield  {author} {\bibinfo {author} {\bibfnamefont {Achille}\ \bibnamefont
  {Papapetrou}},\ }\bibfield  {title} {\enquote {\bibinfo {title} {{Spinning
  test particles in general relativity. 1.}}}\ }\href {\doibase
  10.1098/rspa.1951.0200} {\bibfield  {journal} {\bibinfo  {journal} {Proc.
  Roy. Soc. Lond.}\ }\textbf {\bibinfo {volume} {A209}},\ \bibinfo {pages}
  {248--258} (\bibinfo {year} {1951})}\BibitemShut {NoStop}%
\bibitem [{\citenamefont {Dixon}(1964)}]{Dixon1964}%
  \BibitemOpen
  \bibfield  {author} {\bibinfo {author} {\bibfnamefont {W.~G.}\ \bibnamefont
  {Dixon}},\ }\bibfield  {title} {\enquote {\bibinfo {title} {A covariant
  multipole formalism for extended test bodies in general relativity},}\ }\href
  {\doibase 10.1007/BF02734579} {\bibfield  {journal} {\bibinfo  {journal} {Il
  Nuovo Cimento (1955-1965)}\ }\textbf {\bibinfo {volume} {34}},\ \bibinfo
  {pages} {317--339} (\bibinfo {year} {1964})}\BibitemShut {NoStop}%
\bibitem [{\citenamefont {Dixon}(1970)}]{Dixon:1970zza}%
  \BibitemOpen
  \bibfield  {author} {\bibinfo {author} {\bibfnamefont {W.~G.}\ \bibnamefont
  {Dixon}},\ }\bibfield  {title} {\enquote {\bibinfo {title} {{Dynamics of
  extended bodies in general relativity. I. Momentum and angular momentum}},}\
  }\href {\doibase 10.1098/rspa.1970.0020} {\bibfield  {journal} {\bibinfo
  {journal} {Proc. Roy. Soc. Lond.}\ }\textbf {\bibinfo {volume} {A314}},\
  \bibinfo {pages} {499--527} (\bibinfo {year} {1970})}\BibitemShut {NoStop}%
\bibitem [{\citenamefont {Barausse}\ \emph {et~al.}(2009)\citenamefont
  {Barausse}, \citenamefont {Racine},\ and\ \citenamefont
  {Buonanno}}]{Barausse:2009aa}%
  \BibitemOpen
  \bibfield  {author} {\bibinfo {author} {\bibfnamefont {Enrico}\ \bibnamefont
  {Barausse}}, \bibinfo {author} {\bibfnamefont {Etienne}\ \bibnamefont
  {Racine}}, \ and\ \bibinfo {author} {\bibfnamefont {Alessandra}\ \bibnamefont
  {Buonanno}},\ }\bibfield  {title} {\enquote {\bibinfo {title} {{Hamiltonian
  of a spinning test-particle in curved spacetime}},}\ }\href {\doibase
  10.1103/PhysRevD.85.069904, 10.1103/PhysRevD.80.104025} {\bibfield  {journal}
  {\bibinfo  {journal} {Phys. Rev.}\ }\textbf {\bibinfo {volume} {D80}},\
  \bibinfo {pages} {104025} (\bibinfo {year} {2009})},\ \bibinfo {note}
  {[Erratum: Phys. Rev.D85,069904(2012)]},\ \Eprint
  {http://arxiv.org/abs/0907.4745} {arXiv:0907.4745 [gr-qc]} \BibitemShut
  {NoStop}%
\bibitem [{\citenamefont {Kunst}\ \emph {et~al.}(2016)\citenamefont {Kunst},
  \citenamefont {Ledvinka}, \citenamefont {Lukes-Gerakopoulos},\ and\
  \citenamefont {Seyrich}}]{Kunst:2015tla}%
  \BibitemOpen
  \bibfield  {author} {\bibinfo {author} {\bibfnamefont {Daniela}\ \bibnamefont
  {Kunst}}, \bibinfo {author} {\bibfnamefont {Tomáš}\ \bibnamefont
  {Ledvinka}}, \bibinfo {author} {\bibfnamefont {Georgios}\ \bibnamefont
  {Lukes-Gerakopoulos}}, \ and\ \bibinfo {author} {\bibfnamefont {Jonathan}\
  \bibnamefont {Seyrich}},\ }\bibfield  {title} {\enquote {\bibinfo {title}
  {{Comparing Hamiltonians of a spinning test particle for different tetrad
  fields}},}\ }\href {\doibase 10.1103/PhysRevD.93.044004} {\bibfield
  {journal} {\bibinfo  {journal} {Phys. Rev.}\ }\textbf {\bibinfo {volume}
  {D93}},\ \bibinfo {pages} {044004} (\bibinfo {year} {2016})},\ \Eprint
  {http://arxiv.org/abs/1506.01473} {arXiv:1506.01473 [gr-qc]} \BibitemShut
  {NoStop}%
\bibitem [{\citenamefont {Cho}(1998)}]{Cho:1997vx}%
  \BibitemOpen
  \bibfield  {author} {\bibinfo {author} {\bibfnamefont {H.~T.}\ \bibnamefont
  {Cho}},\ }\bibfield  {title} {\enquote {\bibinfo {title} {{PostNewtonian
  approximation for spinning particles}},}\ }\href {\doibase
  10.1088/0264-9381/15/8/022} {\bibfield  {journal} {\bibinfo  {journal}
  {Class. Quant. Grav.}\ }\textbf {\bibinfo {volume} {15}},\ \bibinfo {pages}
  {2465--2478} (\bibinfo {year} {1998})},\ \Eprint
  {http://arxiv.org/abs/gr-qc/9703071} {arXiv:gr-qc/9703071 [gr-qc]}
  \BibitemShut {NoStop}%
\bibitem [{\citenamefont {Porto}(2006)}]{Porto:2005ac}%
  \BibitemOpen
  \bibfield  {author} {\bibinfo {author} {\bibfnamefont {Rafael~A.}\
  \bibnamefont {Porto}},\ }\bibfield  {title} {\enquote {\bibinfo {title}
  {{Post-Newtonian corrections to the motion of spinning bodies in NRGR}},}\
  }\href {\doibase 10.1103/PhysRevD.73.104031} {\bibfield  {journal} {\bibinfo
  {journal} {Phys. Rev.}\ }\textbf {\bibinfo {volume} {D73}},\ \bibinfo {pages}
  {104031} (\bibinfo {year} {2006})},\ \Eprint
  {http://arxiv.org/abs/gr-qc/0511061} {arXiv:gr-qc/0511061 [gr-qc]}
  \BibitemShut {NoStop}%
\bibitem [{\citenamefont {Apostolatos}(1996)}]{Apostolatos}%
  \BibitemOpen
  \bibfield  {author} {\bibinfo {author} {\bibfnamefont {Theocharis~A}\
  \bibnamefont {Apostolatos}},\ }\bibfield  {title} {\enquote {\bibinfo {title}
  {A spinning test body in the strong field of a schwarzschild black hole},}\
  }\href {http://stacks.iop.org/0264-9381/13/i=5/a=005} {\bibfield  {journal}
  {\bibinfo  {journal} {Classical and Quantum Gravity}\ }\textbf {\bibinfo
  {volume} {13}},\ \bibinfo {pages} {799} (\bibinfo {year} {1996})}\BibitemShut
  {NoStop}%
\bibitem [{\citenamefont {Chicone}\ \emph {et~al.}(2005)\citenamefont
  {Chicone}, \citenamefont {Mashhoon},\ and\ \citenamefont
  {Punsly}}]{Chicone:2005jj}%
  \BibitemOpen
  \bibfield  {author} {\bibinfo {author} {\bibfnamefont {C.}~\bibnamefont
  {Chicone}}, \bibinfo {author} {\bibfnamefont {B.}~\bibnamefont {Mashhoon}}, \
  and\ \bibinfo {author} {\bibfnamefont {B.}~\bibnamefont {Punsly}},\
  }\bibfield  {title} {\enquote {\bibinfo {title} {{Relativistic motion of
  spinning particles in a gravitational field}},}\ }\href {\doibase
  10.1016/j.physleta.2005.05.072} {\bibfield  {journal} {\bibinfo  {journal}
  {Phys. Lett.}\ }\textbf {\bibinfo {volume} {A343}},\ \bibinfo {pages} {1--7}
  (\bibinfo {year} {2005})},\ \Eprint {http://arxiv.org/abs/gr-qc/0504146}
  {arXiv:gr-qc/0504146 [gr-qc]} \BibitemShut {NoStop}%
\bibitem [{\citenamefont {Steinhoff}(2011)}]{Steinhoff:2010zz}%
  \BibitemOpen
  \bibfield  {author} {\bibinfo {author} {\bibfnamefont {Jan}\ \bibnamefont
  {Steinhoff}},\ }\bibfield  {title} {\enquote {\bibinfo {title} {{Canonical
  formulation of spin in general relativity}},}\ }\href {\doibase
  10.1002/andp.201000178} {\bibfield  {journal} {\bibinfo  {journal} {Annalen
  Phys.}\ }\textbf {\bibinfo {volume} {523}},\ \bibinfo {pages} {296--353}
  (\bibinfo {year} {2011})},\ \Eprint {http://arxiv.org/abs/1106.4203}
  {arXiv:1106.4203 [gr-qc]} \BibitemShut {NoStop}%
\bibitem [{\citenamefont {Dixon}(1973)}]{Dixon1973}%
  \BibitemOpen
  \bibfield  {author} {\bibinfo {author} {\bibfnamefont {W.~G.}\ \bibnamefont
  {Dixon}},\ }\bibfield  {title} {\enquote {\bibinfo {title} {The definition of
  multipole moments for extended bodies},}\ }\href {\doibase
  10.1007/BF02412488} {\bibfield  {journal} {\bibinfo  {journal} {General
  Relativity and Gravitation}\ }\textbf {\bibinfo {volume} {4}},\ \bibinfo
  {pages} {199--209} (\bibinfo {year} {1973})}\BibitemShut {NoStop}%
\bibitem [{\citenamefont {Mashhoon}\ and\ \citenamefont
  {Singh}(2006)}]{Mashhoon:2006fj}%
  \BibitemOpen
  \bibfield  {author} {\bibinfo {author} {\bibfnamefont {Bahram}\ \bibnamefont
  {Mashhoon}}\ and\ \bibinfo {author} {\bibfnamefont {Dinesh}\ \bibnamefont
  {Singh}},\ }\bibfield  {title} {\enquote {\bibinfo {title} {{Dynamics of
  Extended Spinning Masses in a Gravitational Field}},}\ }\href {\doibase
  10.1103/PhysRevD.74.124006} {\bibfield  {journal} {\bibinfo  {journal} {Phys.
  Rev.}\ }\textbf {\bibinfo {volume} {D74}},\ \bibinfo {pages} {124006}
  (\bibinfo {year} {2006})},\ \Eprint {http://arxiv.org/abs/astro-ph/0608278}
  {arXiv:astro-ph/0608278 [astro-ph]} \BibitemShut {NoStop}%
\bibitem [{\citenamefont {Vines}\ \emph {et~al.}(2016)\citenamefont {Vines},
  \citenamefont {Kunst}, \citenamefont {Steinhoff},\ and\ \citenamefont
  {Hinderer}}]{Vines:2016unv}%
  \BibitemOpen
  \bibfield  {author} {\bibinfo {author} {\bibfnamefont {Justin}\ \bibnamefont
  {Vines}}, \bibinfo {author} {\bibfnamefont {Daniela}\ \bibnamefont {Kunst}},
  \bibinfo {author} {\bibfnamefont {Jan}\ \bibnamefont {Steinhoff}}, \ and\
  \bibinfo {author} {\bibfnamefont {Tanja}\ \bibnamefont {Hinderer}},\
  }\bibfield  {title} {\enquote {\bibinfo {title} {{Canonical Hamiltonian for
  an extended test body in curved spacetime: To quadratic order in spin}},}\
  }\href {\doibase 10.1103/PhysRevD.93.103008} {\bibfield  {journal} {\bibinfo
  {journal} {Phys. Rev.}\ }\textbf {\bibinfo {volume} {D93}},\ \bibinfo {pages}
  {103008} (\bibinfo {year} {2016})},\ \Eprint
  {http://arxiv.org/abs/1601.07529} {arXiv:1601.07529 [gr-qc]} \BibitemShut
  {NoStop}%
\bibitem [{\citenamefont {Roshan}(2013)}]{Roshan:2012qy}%
  \BibitemOpen
  \bibfield  {author} {\bibinfo {author} {\bibfnamefont {Mahmood}\ \bibnamefont
  {Roshan}},\ }\bibfield  {title} {\enquote {\bibinfo {title} {{Test particle
  motion in modified gravity theories}},}\ }\href {\doibase
  10.1103/PhysRevD.87.044005} {\bibfield  {journal} {\bibinfo  {journal} {Phys.
  Rev.}\ }\textbf {\bibinfo {volume} {D87}},\ \bibinfo {pages} {044005}
  (\bibinfo {year} {2013})},\ \Eprint {http://arxiv.org/abs/1210.3136}
  {arXiv:1210.3136 [gr-qc]} \BibitemShut {NoStop}%
\bibitem [{\citenamefont {Plyatsko}\ and\ \citenamefont
  {Stefanyshyn}(2008{\natexlab{a}})}]{Plyatsko:2008rd}%
  \BibitemOpen
  \bibfield  {author} {\bibinfo {author} {\bibfnamefont {Roman}\ \bibnamefont
  {Plyatsko}}\ and\ \bibinfo {author} {\bibfnamefont {Oleksandr}\ \bibnamefont
  {Stefanyshyn}},\ }\bibfield  {title} {\enquote {\bibinfo {title} {{Mathisson
  Equations: Non-Oscillatory Solutions in a Schwarzschild Field}},}\
  }\href@noop {} {\bibfield  {journal} {\bibinfo  {journal} {Acta Phys.
  Polon.}\ }\textbf {\bibinfo {volume} {B39}},\ \bibinfo {pages} {23} (\bibinfo
  {year} {2008}{\natexlab{a}})},\ \Eprint {http://arxiv.org/abs/0802.0652}
  {arXiv:0802.0652 [gr-qc]} \BibitemShut {NoStop}%
\bibitem [{\citenamefont {Plyatsko}\ and\ \citenamefont
  {Stefanyshyn}(2008{\natexlab{b}})}]{Plyatsko:2008wh}%
  \BibitemOpen
  \bibfield  {author} {\bibinfo {author} {\bibfnamefont {Roman}\ \bibnamefont
  {Plyatsko}}\ and\ \bibinfo {author} {\bibfnamefont {Oleksandr}\ \bibnamefont
  {Stefanyshyn}},\ }\bibfield  {title} {\enquote {\bibinfo {title} {{On common
  solutions of Mathisson equations under different conditions}},}\ }\href@noop
  {} {\  (\bibinfo {year} {2008}{\natexlab{b}})},\ \Eprint
  {http://arxiv.org/abs/0803.0121} {arXiv:0803.0121 [gr-qc]} \BibitemShut
  {NoStop}%
\bibitem [{\citenamefont {Plyatsko}(2005)}]{Plyatsko:2005bh}%
  \BibitemOpen
  \bibfield  {author} {\bibinfo {author} {\bibfnamefont {Roman}\ \bibnamefont
  {Plyatsko}},\ }\bibfield  {title} {\enquote {\bibinfo {title}
  {{Ultrarelativistic circular orbits of spinning particles in a Schwarzschild
  field}},}\ }\href {\doibase 10.1088/0264-9381/22/9/004} {\bibfield  {journal}
  {\bibinfo  {journal} {Class. Quant. Grav.}\ }\textbf {\bibinfo {volume}
  {22}},\ \bibinfo {pages} {1545--1551} (\bibinfo {year} {2005})},\ \Eprint
  {http://arxiv.org/abs/gr-qc/0507023} {arXiv:gr-qc/0507023 [gr-qc]}
  \BibitemShut {NoStop}%
\bibitem [{\citenamefont {Plyatsko}\ and\ \citenamefont
  {Fenyk}(2016)}]{Plyatsko:2016bee}%
  \BibitemOpen
  \bibfield  {author} {\bibinfo {author} {\bibfnamefont {Roman}\ \bibnamefont
  {Plyatsko}}\ and\ \bibinfo {author} {\bibfnamefont {Mykola}\ \bibnamefont
  {Fenyk}},\ }\bibfield  {title} {\enquote {\bibinfo {title} {{Antigravity:
  Spin-gravity coupling in action}},}\ }\href {\doibase
  10.1103/PhysRevD.94.044047} {\bibfield  {journal} {\bibinfo  {journal} {Phys.
  Rev.}\ }\textbf {\bibinfo {volume} {D94}},\ \bibinfo {pages} {044047}
  (\bibinfo {year} {2016})},\ \Eprint {http://arxiv.org/abs/1610.01545}
  {arXiv:1610.01545 [gr-qc]} \BibitemShut {NoStop}%
\bibitem [{\citenamefont {Suzuki}\ and\ \citenamefont
  {Maeda}(1997)}]{Suzuki:1996gm}%
  \BibitemOpen
  \bibfield  {author} {\bibinfo {author} {\bibfnamefont {Shingo}\ \bibnamefont
  {Suzuki}}\ and\ \bibinfo {author} {\bibfnamefont {Kei-ichi}\ \bibnamefont
  {Maeda}},\ }\bibfield  {title} {\enquote {\bibinfo {title} {{Chaos in
  Schwarzschild space-time: The motion of a spinning particle}},}\ }\href
  {\doibase 10.1103/PhysRevD.55.4848} {\bibfield  {journal} {\bibinfo
  {journal} {Phys. Rev.}\ }\textbf {\bibinfo {volume} {D55}},\ \bibinfo {pages}
  {4848--4859} (\bibinfo {year} {1997})},\ \Eprint
  {http://arxiv.org/abs/gr-qc/9604020} {arXiv:gr-qc/9604020 [gr-qc]}
  \BibitemShut {NoStop}%
\bibitem [{\citenamefont {Rietdijk}\ and\ \citenamefont {van
  Holten}(1993)}]{0264-9381-10-3-017}%
  \BibitemOpen
  \bibfield  {author} {\bibinfo {author} {\bibfnamefont {R~H}\ \bibnamefont
  {Rietdijk}}\ and\ \bibinfo {author} {\bibfnamefont {J~W}\ \bibnamefont {van
  Holten}},\ }\bibfield  {title} {\enquote {\bibinfo {title} {Spinning
  particles in schwarzschild spacetime},}\ }\href
  {http://stacks.iop.org/0264-9381/10/i=3/a=017} {\bibfield  {journal}
  {\bibinfo  {journal} {Classical and Quantum Gravity}\ }\textbf {\bibinfo
  {volume} {10}},\ \bibinfo {pages} {575} (\bibinfo {year} {1993})}\BibitemShut
  {NoStop}%
\bibitem [{\citenamefont {Bini}\ \emph {et~al.}(2005)\citenamefont {Bini},
  \citenamefont {de~Felice}, \citenamefont {Geralico},\ and\ \citenamefont
  {Jantzen}}]{Bini:2005nt}%
  \BibitemOpen
  \bibfield  {author} {\bibinfo {author} {\bibfnamefont {Donato}\ \bibnamefont
  {Bini}}, \bibinfo {author} {\bibfnamefont {Fernando}\ \bibnamefont
  {de~Felice}}, \bibinfo {author} {\bibfnamefont {Andrea}\ \bibnamefont
  {Geralico}}, \ and\ \bibinfo {author} {\bibfnamefont {Robert~T.}\
  \bibnamefont {Jantzen}},\ }\bibfield  {title} {\enquote {\bibinfo {title}
  {{Spin precession in the Schwarzschild spacetime: Circular orbits}},}\ }\href
  {\doibase 10.1088/0264-9381/22/14/007} {\bibfield  {journal} {\bibinfo
  {journal} {Class. Quant. Grav.}\ }\textbf {\bibinfo {volume} {22}},\ \bibinfo
  {pages} {2947--2970} (\bibinfo {year} {2005})},\ \Eprint
  {http://arxiv.org/abs/gr-qc/0506017} {arXiv:gr-qc/0506017 [gr-qc]}
  \BibitemShut {NoStop}%
\bibitem [{\citenamefont {Plyatsko}\ \emph {et~al.}(2011)\citenamefont
  {Plyatsko}, \citenamefont {Stefanyshyn},\ and\ \citenamefont
  {Fenyk}}]{Plyatsko:2011gf}%
  \BibitemOpen
  \bibfield  {author} {\bibinfo {author} {\bibfnamefont {Roman}\ \bibnamefont
  {Plyatsko}}, \bibinfo {author} {\bibfnamefont {Oleksandr}\ \bibnamefont
  {Stefanyshyn}}, \ and\ \bibinfo {author} {\bibfnamefont {Mykola}\
  \bibnamefont {Fenyk}},\ }\bibfield  {title} {\enquote {\bibinfo {title}
  {{Mathisson-Papapetrou-Dixon equations in the Schwarzschild and Kerr
  backgrounds}},}\ }\href {\doibase 10.1088/0264-9381/28/19/195025} {\bibfield
  {journal} {\bibinfo  {journal} {Class. Quant. Grav.}\ }\textbf {\bibinfo
  {volume} {28}},\ \bibinfo {pages} {195025} (\bibinfo {year} {2011})},\
  \Eprint {http://arxiv.org/abs/1110.1967} {arXiv:1110.1967 [gr-qc]}
  \BibitemShut {NoStop}%
\bibitem [{\citenamefont {Plyatsko}\ and\ \citenamefont
  {Fenyk}(2013)}]{Plyatsko:2013xza}%
  \BibitemOpen
  \bibfield  {author} {\bibinfo {author} {\bibfnamefont {Roman}\ \bibnamefont
  {Plyatsko}}\ and\ \bibinfo {author} {\bibfnamefont {Mykola}\ \bibnamefont
  {Fenyk}},\ }\bibfield  {title} {\enquote {\bibinfo {title} {{Highly
  relativistic circular orbits of spinning particle in the Kerr field}},}\
  }\href {\doibase 10.1103/PhysRevD.87.044019} {\bibfield  {journal} {\bibinfo
  {journal} {Phys. Rev.}\ }\textbf {\bibinfo {volume} {D87}},\ \bibinfo {pages}
  {044019} (\bibinfo {year} {2013})},\ \Eprint {http://arxiv.org/abs/1303.4707}
  {arXiv:1303.4707 [gr-qc]} \BibitemShut {NoStop}%
\bibitem [{\citenamefont {Lukes-Gerakopoulos}\ \emph
  {et~al.}(2016)\citenamefont {Lukes-Gerakopoulos}, \citenamefont {Katsanikas},
  \citenamefont {Patsis},\ and\ \citenamefont
  {Seyrich}}]{Lukes-Gerakopoulos:2016bup}%
  \BibitemOpen
  \bibfield  {author} {\bibinfo {author} {\bibfnamefont {Georgios}\
  \bibnamefont {Lukes-Gerakopoulos}}, \bibinfo {author} {\bibfnamefont
  {Matthaios}\ \bibnamefont {Katsanikas}}, \bibinfo {author} {\bibfnamefont
  {Panos~A.}\ \bibnamefont {Patsis}}, \ and\ \bibinfo {author} {\bibfnamefont
  {Jonathan}\ \bibnamefont {Seyrich}},\ }\bibfield  {title} {\enquote {\bibinfo
  {title} {{The dynamics of a spinning particle in a linear in spin Hamiltonian
  approximation}},}\ }\href {\doibase 10.1103/PhysRevD.94.024024} {\bibfield
  {journal} {\bibinfo  {journal} {Phys. Rev.}\ }\textbf {\bibinfo {volume}
  {D94}},\ \bibinfo {pages} {024024} (\bibinfo {year} {2016})},\ \Eprint
  {http://arxiv.org/abs/1606.09171} {arXiv:1606.09171 [gr-qc]} \BibitemShut
  {NoStop}%
\bibitem [{\citenamefont {Bini}\ \emph {et~al.}(2006)\citenamefont {Bini},
  \citenamefont {Geralico}, \citenamefont {Jantzen},\ and\ \citenamefont
  {de~Felice}}]{Bini:2006pc}%
  \BibitemOpen
  \bibfield  {author} {\bibinfo {author} {\bibfnamefont {D.}~\bibnamefont
  {Bini}}, \bibinfo {author} {\bibfnamefont {A.}~\bibnamefont {Geralico}},
  \bibinfo {author} {\bibfnamefont {R.~T.}\ \bibnamefont {Jantzen}}, \ and\
  \bibinfo {author} {\bibfnamefont {F.}~\bibnamefont {de~Felice}},\ }\bibfield
  {title} {\enquote {\bibinfo {title} {{Spin precession along circular orbits
  in the Kerr spacetime: the Frenet-Serret description}},}\ }\href {\doibase
  10.1088/0264-9381/23/10/003} {\bibfield  {journal} {\bibinfo  {journal}
  {Class. Quant. Grav.}\ }\textbf {\bibinfo {volume} {23}},\ \bibinfo {pages}
  {3287--3304} (\bibinfo {year} {2006})},\ \Eprint
  {http://arxiv.org/abs/1408.4278} {arXiv:1408.4278 [gr-qc]} \BibitemShut
  {NoStop}%
\bibitem [{\citenamefont {Hartl}(2003{\natexlab{b}})}]{Hartl:2003da}%
  \BibitemOpen
  \bibfield  {author} {\bibinfo {author} {\bibfnamefont {Michael~D.}\
  \bibnamefont {Hartl}},\ }\bibfield  {title} {\enquote {\bibinfo {title} {{A
  Survey of spinning test particle orbits in Kerr space-time}},}\ }\href
  {\doibase 10.1103/PhysRevD.67.104023} {\bibfield  {journal} {\bibinfo
  {journal} {Phys. Rev.}\ }\textbf {\bibinfo {volume} {D67}},\ \bibinfo {pages}
  {104023} (\bibinfo {year} {2003}{\natexlab{b}})},\ \Eprint
  {http://arxiv.org/abs/gr-qc/0302103} {arXiv:gr-qc/0302103 [gr-qc]}
  \BibitemShut {NoStop}%
\bibitem [{\citenamefont {Suzuki}\ and\ \citenamefont
  {Maeda}(1998)}]{Suzuki:1997by}%
  \BibitemOpen
  \bibfield  {author} {\bibinfo {author} {\bibfnamefont {Shingo}\ \bibnamefont
  {Suzuki}}\ and\ \bibinfo {author} {\bibfnamefont {Kei-ichi}\ \bibnamefont
  {Maeda}},\ }\bibfield  {title} {\enquote {\bibinfo {title} {{Innermost stable
  circular orbit of a spinning particle in Kerr space-time}},}\ }\href
  {\doibase 10.1103/PhysRevD.58.023005} {\bibfield  {journal} {\bibinfo
  {journal} {Phys. Rev.}\ }\textbf {\bibinfo {volume} {D58}},\ \bibinfo {pages}
  {023005} (\bibinfo {year} {1998})},\ \Eprint
  {http://arxiv.org/abs/gr-qc/9712095} {arXiv:gr-qc/9712095 [gr-qc]}
  \BibitemShut {NoStop}%
\bibitem [{\citenamefont {Han}(2008)}]{Han:2008zzf}%
  \BibitemOpen
  \bibfield  {author} {\bibinfo {author} {\bibfnamefont {Wenbiao}\ \bibnamefont
  {Han}},\ }\bibfield  {title} {\enquote {\bibinfo {title} {{Chaos and dynamics
  of spinning particles in Kerr spacetime}},}\ }\href {\doibase
  10.1007/s10714-007-0598-9} {\bibfield  {journal} {\bibinfo  {journal} {Gen.
  Rel. Grav.}\ }\textbf {\bibinfo {volume} {40}},\ \bibinfo {pages}
  {1831--1847} (\bibinfo {year} {2008})},\ \Eprint
  {http://arxiv.org/abs/1006.2229} {arXiv:1006.2229 [gr-qc]} \BibitemShut
  {NoStop}%
\bibitem [{\citenamefont {Hackmann}\ \emph {et~al.}(2014)\citenamefont
  {Hackmann}, \citenamefont {Lämmerzahl}, \citenamefont {Obukhov},
  \citenamefont {Puetzfeld},\ and\ \citenamefont
  {Schaffer}}]{Hackmann:2014tga}%
  \BibitemOpen
  \bibfield  {author} {\bibinfo {author} {\bibfnamefont {Eva}\ \bibnamefont
  {Hackmann}}, \bibinfo {author} {\bibfnamefont {Claus}\ \bibnamefont
  {Lämmerzahl}}, \bibinfo {author} {\bibfnamefont {Yuri~N.}\ \bibnamefont
  {Obukhov}}, \bibinfo {author} {\bibfnamefont {Dirk}\ \bibnamefont
  {Puetzfeld}}, \ and\ \bibinfo {author} {\bibfnamefont {Isabell}\ \bibnamefont
  {Schaffer}},\ }\bibfield  {title} {\enquote {\bibinfo {title} {{Motion of
  spinning test bodies in Kerr spacetime}},}\ }\href {\doibase
  10.1103/PhysRevD.90.064035} {\bibfield  {journal} {\bibinfo  {journal} {Phys.
  Rev.}\ }\textbf {\bibinfo {volume} {D90}},\ \bibinfo {pages} {064035}
  (\bibinfo {year} {2014})},\ \Eprint {http://arxiv.org/abs/1408.1773}
  {arXiv:1408.1773 [gr-qc]} \BibitemShut {NoStop}%
\bibitem [{\citenamefont {Mohseni}(2010)}]{Mohseni:2010rm}%
  \BibitemOpen
  \bibfield  {author} {\bibinfo {author} {\bibfnamefont {Morteza}\ \bibnamefont
  {Mohseni}},\ }\bibfield  {title} {\enquote {\bibinfo {title} {{Stability of
  circular orbits of spinning particles in Schwarzschild-like space-times}},}\
  }\href {\doibase 10.1007/s10714-010-0995-3} {\bibfield  {journal} {\bibinfo
  {journal} {Gen. Rel. Grav.}\ }\textbf {\bibinfo {volume} {42}},\ \bibinfo
  {pages} {2477--2490} (\bibinfo {year} {2010})},\ \Eprint
  {http://arxiv.org/abs/1005.3110} {arXiv:1005.3110 [gr-qc]} \BibitemShut
  {NoStop}%
\bibitem [{\citenamefont {Mortazavimanesh}\ and\ \citenamefont
  {Mohseni}(2009)}]{Mortazavimanesh:2009rm}%
  \BibitemOpen
  \bibfield  {author} {\bibinfo {author} {\bibfnamefont {M.}~\bibnamefont
  {Mortazavimanesh}}\ and\ \bibinfo {author} {\bibfnamefont {Morteza}\
  \bibnamefont {Mohseni}},\ }\bibfield  {title} {\enquote {\bibinfo {title}
  {{Spinning particles in Schwarzschild-de Sitter space-time}},}\ }\href
  {\doibase 10.1007/s10714-009-0798-6} {\bibfield  {journal} {\bibinfo
  {journal} {Gen. Rel. Grav.}\ }\textbf {\bibinfo {volume} {41}},\ \bibinfo
  {pages} {2697--2706} (\bibinfo {year} {2009})},\ \Eprint
  {http://arxiv.org/abs/0904.1263} {arXiv:0904.1263 [gr-qc]} \BibitemShut
  {NoStop}%
\bibitem [{\citenamefont {Obukhov}\ and\ \citenamefont
  {Puetzfeld}(2011)}]{Obukhov:2010kn}%
  \BibitemOpen
  \bibfield  {author} {\bibinfo {author} {\bibfnamefont {Yuri~N.}\ \bibnamefont
  {Obukhov}}\ and\ \bibinfo {author} {\bibfnamefont {Dirk}\ \bibnamefont
  {Puetzfeld}},\ }\bibfield  {title} {\enquote {\bibinfo {title} {{Dynamics of
  test bodies with spin in de Sitter spacetime}},}\ }\href {\doibase
  10.1103/PhysRevD.83.044024} {\bibfield  {journal} {\bibinfo  {journal} {Phys.
  Rev.}\ }\textbf {\bibinfo {volume} {D83}},\ \bibinfo {pages} {044024}
  (\bibinfo {year} {2011})},\ \Eprint {http://arxiv.org/abs/1010.1451}
  {arXiv:1010.1451 [gr-qc]} \BibitemShut {NoStop}%
\bibitem [{\citenamefont {Stuchlik}\ and\ \citenamefont
  {Slany}(2004)}]{Stuchlik:2003dt}%
  \BibitemOpen
  \bibfield  {author} {\bibinfo {author} {\bibfnamefont {Zdenek}\ \bibnamefont
  {Stuchlik}}\ and\ \bibinfo {author} {\bibfnamefont {Petr}\ \bibnamefont
  {Slany}},\ }\bibfield  {title} {\enquote {\bibinfo {title} {{Equatorial
  circular orbits in the Kerr-de Sitter space-times}},}\ }\href {\doibase
  10.1103/PhysRevD.69.064001} {\bibfield  {journal} {\bibinfo  {journal} {Phys.
  Rev.}\ }\textbf {\bibinfo {volume} {D69}},\ \bibinfo {pages} {064001}
  (\bibinfo {year} {2004})},\ \Eprint {http://arxiv.org/abs/gr-qc/0307049}
  {arXiv:gr-qc/0307049 [gr-qc]} \BibitemShut {NoStop}%
\bibitem [{\citenamefont {Zalaquett}\ \emph {et~al.}(2014)\citenamefont
  {Zalaquett}, \citenamefont {Hojman},\ and\ \citenamefont
  {Asenjo}}]{Zalaquett:2014eia}%
  \BibitemOpen
  \bibfield  {author} {\bibinfo {author} {\bibfnamefont {Nicolas}\ \bibnamefont
  {Zalaquett}}, \bibinfo {author} {\bibfnamefont {Sergio~A.}\ \bibnamefont
  {Hojman}}, \ and\ \bibinfo {author} {\bibfnamefont {Felipe~A.}\ \bibnamefont
  {Asenjo}},\ }\bibfield  {title} {\enquote {\bibinfo {title} {{Spinning
  massive test particles in cosmological and general static spherically
  symmetric spacetimes}},}\ }\href {\doibase 10.1088/0264-9381/31/8/085011}
  {\bibfield  {journal} {\bibinfo  {journal} {Class. Quant. Grav.}\ }\textbf
  {\bibinfo {volume} {31}},\ \bibinfo {pages} {085011} (\bibinfo {year}
  {2014})},\ \Eprint {http://arxiv.org/abs/1308.4435} {arXiv:1308.4435 [gr-qc]}
  \BibitemShut {NoStop}%
\bibitem [{\citenamefont {Hojman}\ and\ \citenamefont
  {Hojman}(1977)}]{Hojman:1976kn}%
  \BibitemOpen
  \bibfield  {author} {\bibinfo {author} {\bibfnamefont {Roberto}\ \bibnamefont
  {Hojman}}\ and\ \bibinfo {author} {\bibfnamefont {Sergio}\ \bibnamefont
  {Hojman}},\ }\bibfield  {title} {\enquote {\bibinfo {title} {{Spinning
  Charged Test Particles in a Kerr-Newman Background}},}\ }\href {\doibase
  10.1103/PhysRevD.15.2724} {\bibfield  {journal} {\bibinfo  {journal} {Phys.
  Rev.}\ }\textbf {\bibinfo {volume} {D15}},\ \bibinfo {pages} {2724} (\bibinfo
  {year} {1977})}\BibitemShut {NoStop}%
\bibitem [{\citenamefont {Bini}\ \emph {et~al.}(2000)\citenamefont {Bini},
  \citenamefont {Gemelli},\ and\ \citenamefont {Ruffini}}]{Bini:2000vv}%
  \BibitemOpen
  \bibfield  {author} {\bibinfo {author} {\bibfnamefont {D.}~\bibnamefont
  {Bini}}, \bibinfo {author} {\bibfnamefont {G.}~\bibnamefont {Gemelli}}, \
  and\ \bibinfo {author} {\bibfnamefont {R.}~\bibnamefont {Ruffini}},\
  }\bibfield  {title} {\enquote {\bibinfo {title} {{Spinning test particles in
  general relativity: Nongeodesic motion in the Reissner-Nordstrom
  space-time}},}\ }\href {\doibase 10.1103/PhysRevD.61.064013} {\bibfield
  {journal} {\bibinfo  {journal} {Phys. Rev.}\ }\textbf {\bibinfo {volume}
  {D61}},\ \bibinfo {pages} {064013} (\bibinfo {year} {2000})}\BibitemShut
  {NoStop}%
\bibitem [{\citenamefont {Nomura}\ \emph {et~al.}(1991)\citenamefont {Nomura},
  \citenamefont {Shirafuji},\ and\ \citenamefont {Hayashi}}]{Nomura:1991yx}%
  \BibitemOpen
  \bibfield  {author} {\bibinfo {author} {\bibfnamefont {Koichi}\ \bibnamefont
  {Nomura}}, \bibinfo {author} {\bibfnamefont {Takeshi}\ \bibnamefont
  {Shirafuji}}, \ and\ \bibinfo {author} {\bibfnamefont {Kenji}\ \bibnamefont
  {Hayashi}},\ }\bibfield  {title} {\enquote {\bibinfo {title} {{Spinning test
  particles in space-time with torsion}},}\ }\href {\doibase
  10.1143/PTP.86.1239} {\bibfield  {journal} {\bibinfo  {journal} {Prog. Theor.
  Phys.}\ }\textbf {\bibinfo {volume} {86}},\ \bibinfo {pages} {1239--1258}
  (\bibinfo {year} {1991})}\BibitemShut {NoStop}%
\bibitem [{\citenamefont {Maity}\ \emph {et~al.}(2005)\citenamefont {Maity},
  \citenamefont {SenGupta},\ and\ \citenamefont {Sur}}]{Maity:2004yk}%
  \BibitemOpen
  \bibfield  {author} {\bibinfo {author} {\bibfnamefont {Debaprasad}\
  \bibnamefont {Maity}}, \bibinfo {author} {\bibfnamefont {Soumitra}\
  \bibnamefont {SenGupta}}, \ and\ \bibinfo {author} {\bibfnamefont {Saurabh}\
  \bibnamefont {Sur}},\ }\bibfield  {title} {\enquote {\bibinfo {title}
  {{Spinning test particle in Kalb-Ramond background}},}\ }\href {\doibase
  10.1140/epjc/s2005-02297-6} {\bibfield  {journal} {\bibinfo  {journal} {Eur.
  Phys. J.}\ }\textbf {\bibinfo {volume} {C42}},\ \bibinfo {pages} {453--460}
  (\bibinfo {year} {2005})},\ \Eprint {http://arxiv.org/abs/hep-th/0409143}
  {arXiv:hep-th/0409143 [hep-th]} \BibitemShut {NoStop}%
\bibitem [{\citenamefont
  {Lukes-Gerakopoulos}(2017{\natexlab{b}})}]{Lukes-Gerakopoulos:2017cru}%
  \BibitemOpen
  \bibfield  {author} {\bibinfo {author} {\bibfnamefont {Georgios}\
  \bibnamefont {Lukes-Gerakopoulos}},\ }\bibfield  {title} {\enquote {\bibinfo
  {title} {{Time parameterizations and spin supplementary conditions of the
  Mathisson-Papapetrou-Dixon equations}},}\ }\href {\doibase
  10.1103/PhysRevD.96.104023} {\bibfield  {journal} {\bibinfo  {journal} {Phys.
  Rev.}\ }\textbf {\bibinfo {volume} {D96}},\ \bibinfo {pages} {104023}
  (\bibinfo {year} {2017}{\natexlab{b}})},\ \Eprint
  {http://arxiv.org/abs/1709.08942} {arXiv:1709.08942 [gr-qc]} \BibitemShut
  {NoStop}%
\bibitem [{\citenamefont {Costa}\ \emph {et~al.}(2012)\citenamefont {Costa},
  \citenamefont {Herdeiro}, \citenamefont {Natario},\ and\ \citenamefont
  {Zilhao}}]{Costa:2011zn}%
  \BibitemOpen
  \bibfield  {author} {\bibinfo {author} {\bibfnamefont {Filipe}\ \bibnamefont
  {Costa}}, \bibinfo {author} {\bibfnamefont {Carlos A.~R.}\ \bibnamefont
  {Herdeiro}}, \bibinfo {author} {\bibfnamefont {Jose}\ \bibnamefont
  {Natario}}, \ and\ \bibinfo {author} {\bibfnamefont {Miguel}\ \bibnamefont
  {Zilhao}},\ }\bibfield  {title} {\enquote {\bibinfo {title} {{Mathisson's
  helical motions for a spinning particle: Are they unphysical?}}}\ }\href
  {\doibase 10.1103/PhysRevD.85.024001} {\bibfield  {journal} {\bibinfo
  {journal} {Phys. Rev.}\ }\textbf {\bibinfo {volume} {D85}},\ \bibinfo {pages}
  {024001} (\bibinfo {year} {2012})},\ \Eprint {http://arxiv.org/abs/1109.1019}
  {arXiv:1109.1019 [gr-qc]} \BibitemShut {NoStop}%
\bibitem [{\citenamefont {Lukes-Gerakopoulos}\ \emph
  {et~al.}(2017)\citenamefont {Lukes-Gerakopoulos}, \citenamefont {Harms},
  \citenamefont {Bernuzzi},\ and\ \citenamefont
  {Nagar}}]{Lukes-Gerakopoulos:2017vkj}%
  \BibitemOpen
  \bibfield  {author} {\bibinfo {author} {\bibfnamefont {Georgios}\
  \bibnamefont {Lukes-Gerakopoulos}}, \bibinfo {author} {\bibfnamefont {Enno}\
  \bibnamefont {Harms}}, \bibinfo {author} {\bibfnamefont {Sebastiano}\
  \bibnamefont {Bernuzzi}}, \ and\ \bibinfo {author} {\bibfnamefont
  {Alessandro}\ \bibnamefont {Nagar}},\ }\bibfield  {title} {\enquote {\bibinfo
  {title} {{Spinning test-body orbiting around a Kerr black hole: circular
  dynamics and gravitational-wave fluxes}},}\ }\href {\doibase
  10.1103/PhysRevD.96.064051} {\bibfield  {journal} {\bibinfo  {journal} {Phys.
  Rev.}\ }\textbf {\bibinfo {volume} {D96}},\ \bibinfo {pages} {064051}
  (\bibinfo {year} {2017})},\ \Eprint {http://arxiv.org/abs/1707.07537}
  {arXiv:1707.07537 [gr-qc]} \BibitemShut {NoStop}%
\bibitem [{\citenamefont {Pirani}(1956)}]{Pirani:1956tn}%
  \BibitemOpen
  \bibfield  {author} {\bibinfo {author} {\bibfnamefont {F.~A.~E.}\
  \bibnamefont {Pirani}},\ }\bibfield  {title} {\enquote {\bibinfo {title} {{On
  the Physical significance of the Riemann tensor}},}\ }\href {\doibase
  10.1007/s10714-009-0787-9} {\bibfield  {journal} {\bibinfo  {journal} {Acta
  Phys. Polon.}\ }\textbf {\bibinfo {volume} {15}},\ \bibinfo {pages}
  {389--405} (\bibinfo {year} {1956})},\ \bibinfo {note} {[Gen. Rel.
  Grav.41,1215(2009)]}\BibitemShut {NoStop}%
\bibitem [{\citenamefont {Lukes-Gerakopoulos}\ \emph
  {et~al.}(2014)\citenamefont {Lukes-Gerakopoulos}, \citenamefont {Seyrich},\
  and\ \citenamefont {Kunst}}]{Lukes-Gerakopoulos:2014dma}%
  \BibitemOpen
  \bibfield  {author} {\bibinfo {author} {\bibfnamefont {Georgios}\
  \bibnamefont {Lukes-Gerakopoulos}}, \bibinfo {author} {\bibfnamefont
  {Jonathan}\ \bibnamefont {Seyrich}}, \ and\ \bibinfo {author} {\bibfnamefont
  {Daniela}\ \bibnamefont {Kunst}},\ }\bibfield  {title} {\enquote {\bibinfo
  {title} {{Investigating spinning test particles: spin supplementary
  conditions and the Hamiltonian formalism}},}\ }\href {\doibase
  10.1103/PhysRevD.90.104019} {\bibfield  {journal} {\bibinfo  {journal} {Phys.
  Rev.}\ }\textbf {\bibinfo {volume} {D90}},\ \bibinfo {pages} {104019}
  (\bibinfo {year} {2014})},\ \Eprint {http://arxiv.org/abs/1409.4314}
  {arXiv:1409.4314 [gr-qc]} \BibitemShut {NoStop}%
\bibitem [{\citenamefont {Tulczyjew}(1959)}]{Tulczyjew}%
  \BibitemOpen
  \bibfield  {author} {\bibinfo {author} {\bibfnamefont {W}~\bibnamefont
  {Tulczyjew}},\ }\bibfield  {title} {\enquote {\bibinfo {title} {{Motion of
  multipole particles in general relativity theory}},}\ }\href@noop {}
  {\bibfield  {journal} {\bibinfo  {journal} {Acta Phys. Polon.}\ }\textbf
  {\bibinfo {volume} {18}},\ \bibinfo {pages} {393} (\bibinfo {year}
  {1959})}\BibitemShut {NoStop}%
\bibitem [{\citenamefont {Bini}\ \emph {et~al.}(2011)\citenamefont {Bini},
  \citenamefont {Geralico},\ and\ \citenamefont {Jantzen}}]{Bini:2014poa}%
  \BibitemOpen
  \bibfield  {author} {\bibinfo {author} {\bibfnamefont {Donato}\ \bibnamefont
  {Bini}}, \bibinfo {author} {\bibfnamefont {Andrea}\ \bibnamefont {Geralico}},
  \ and\ \bibinfo {author} {\bibfnamefont {Robert~T.}\ \bibnamefont
  {Jantzen}},\ }\bibfield  {title} {\enquote {\bibinfo {title} {{Spin-geodesic
  deviations in the Schwarzschild spacetime}},}\ }\href {\doibase
  10.1007/s10714-010-1111-4} {\bibfield  {journal} {\bibinfo  {journal} {Gen.
  Rel. Grav.}\ }\textbf {\bibinfo {volume} {43}},\ \bibinfo {pages} {959}
  (\bibinfo {year} {2011})},\ \Eprint {http://arxiv.org/abs/1408.4946}
  {arXiv:1408.4946 [gr-qc]} \BibitemShut {NoStop}%
\bibitem [{\citenamefont {Asenjo}\ and\ \citenamefont
  {Hojman}(2016)}]{Asenjo:2016uxz}%
  \BibitemOpen
  \bibfield  {author} {\bibinfo {author} {\bibfnamefont {Felipe~A.}\
  \bibnamefont {Asenjo}}\ and\ \bibinfo {author} {\bibfnamefont {Sergio~A.}\
  \bibnamefont {Hojman}},\ }\bibfield  {title} {\enquote {\bibinfo {title}
  {{Exact solutions for the motion of spinning massive particles in conformally
  flat spacetimes}},}\ }\bibfield  {booktitle} {\emph {\bibinfo {booktitle}
  {{Proceedings, 19th Chilean Physics Symposium 2014: Concepción, Chile,
  November 26-28, 2014}}},\ }\href {\doibase 10.1088/1742-6596/720/1/012011}
  {\bibfield  {journal} {\bibinfo  {journal} {J. Phys. Conf. Ser.}\ }\textbf
  {\bibinfo {volume} {720}},\ \bibinfo {pages} {012011} (\bibinfo {year}
  {2016})}\BibitemShut {NoStop}%
\bibitem [{\citenamefont {Kyrian}\ and\ \citenamefont
  {Semerak}(2007)}]{Kyrian:2007zz}%
  \BibitemOpen
  \bibfield  {author} {\bibinfo {author} {\bibfnamefont {K}~\bibnamefont
  {Kyrian}}\ and\ \bibinfo {author} {\bibfnamefont {O}~\bibnamefont
  {Semerak}},\ }\bibfield  {title} {\enquote {\bibinfo {title} {{Spinning test
  particles in a Kerr field}},}\ }\href {\doibase
  10.1111/j.1365-2966.2007.12502.x} {\bibfield  {journal} {\bibinfo  {journal}
  {Mon. Not. Roy. Astron. Soc.}\ }\textbf {\bibinfo {volume} {382}},\ \bibinfo
  {pages} {1922} (\bibinfo {year} {2007})}\BibitemShut {NoStop}%
\bibitem [{\citenamefont {Costa}\ and\ \citenamefont
  {Natário}(2015)}]{Costa:2014nta}%
  \BibitemOpen
  \bibfield  {author} {\bibinfo {author} {\bibfnamefont {L.~Filipe~O.}\
  \bibnamefont {Costa}}\ and\ \bibinfo {author} {\bibfnamefont {José}\
  \bibnamefont {Natário}},\ }\bibfield  {title} {\enquote {\bibinfo {title}
  {{Center of mass, spin supplementary conditions, and the momentum of spinning
  particles}},}\ }\bibfield  {booktitle} {\emph {\bibinfo {booktitle}
  {{Proceedings, 524th WE-Heraeus-Seminar: Equations of Motion in Relativistic
  Gravity (EOM 2013): Bad Honnef, Germany, February 17-23, 2013}}},\ }\href
  {\doibase 10.1007/978-3-319-18335-0_6} {\bibfield  {journal} {\bibinfo
  {journal} {Fund. Theor. Phys.}\ }\textbf {\bibinfo {volume} {179}},\ \bibinfo
  {pages} {215--258} (\bibinfo {year} {2015})},\ \Eprint
  {http://arxiv.org/abs/1410.6443} {arXiv:1410.6443 [gr-qc]} \BibitemShut
  {NoStop}%
\bibitem [{\citenamefont {Costa}\ \emph {et~al.}(2017)\citenamefont {Costa},
  \citenamefont {Lukes-Gerakopoulos},\ and\ \citenamefont
  {Semerák}}]{Costa:2017kdr}%
  \BibitemOpen
  \bibfield  {author} {\bibinfo {author} {\bibfnamefont {L.~Filipe~O.}\
  \bibnamefont {Costa}}, \bibinfo {author} {\bibfnamefont {Georgios}\
  \bibnamefont {Lukes-Gerakopoulos}}, \ and\ \bibinfo {author} {\bibfnamefont
  {Oldřich}\ \bibnamefont {Semerák}},\ }\bibfield  {title} {\enquote
  {\bibinfo {title} {{On spinning particles in general relativity:
  momentum-velocity relation for the Mathisson-Pirani spin condition}},}\
  }\href@noop {} {\  (\bibinfo {year} {2017})},\ \Eprint
  {http://arxiv.org/abs/1712.07281} {arXiv:1712.07281 [gr-qc]} \BibitemShut
  {NoStop}%
\bibitem [{\citenamefont {McVittie}(1933)}]{McVittie:1933zz}%
  \BibitemOpen
  \bibfield  {author} {\bibinfo {author} {\bibfnamefont {G.~C.}\ \bibnamefont
  {McVittie}},\ }\bibfield  {title} {\enquote {\bibinfo {title} {{The
  mass-particle in an expanding universe}},}\ }\href {\doibase
  10.1093/mnras/93.5.325} {\bibfield  {journal} {\bibinfo  {journal} {Mon. Not.
  Roy. Astron. Soc.}\ }\textbf {\bibinfo {volume} {93}},\ \bibinfo {pages}
  {325--339} (\bibinfo {year} {1933})}\BibitemShut {NoStop}%
\bibitem [{\citenamefont {Carrera}\ and\ \citenamefont
  {Giulini}(2010)}]{Carrera:2009ve}%
  \BibitemOpen
  \bibfield  {author} {\bibinfo {author} {\bibfnamefont {Matteo}\ \bibnamefont
  {Carrera}}\ and\ \bibinfo {author} {\bibfnamefont {Domenico}\ \bibnamefont
  {Giulini}},\ }\bibfield  {title} {\enquote {\bibinfo {title} {{On the
  generalization of McVittie's model for an inhomogeneity in a cosmological
  spacetime}},}\ }\href {\doibase 10.1103/PhysRevD.81.043521} {\bibfield
  {journal} {\bibinfo  {journal} {Phys. Rev.}\ }\textbf {\bibinfo {volume}
  {D81}},\ \bibinfo {pages} {043521} (\bibinfo {year} {2010})},\ \Eprint
  {http://arxiv.org/abs/0908.3101} {arXiv:0908.3101 [gr-qc]} \BibitemShut
  {NoStop}%
\bibitem [{\citenamefont {Schutz}(2009)}]{schutz_2009}%
  \BibitemOpen
  \bibfield  {author} {\bibinfo {author} {\bibfnamefont {Bernard}\ \bibnamefont
  {Schutz}},\ }\href {\doibase 10.1017/CBO9780511984181} {\emph {\bibinfo
  {title} {A First Course in General Relativity}}},\ \bibinfo {edition} {2nd}\
  ed.\ (\bibinfo  {publisher} {Cambridge University Press},\ \bibinfo {year}
  {2009})\BibitemShut {NoStop}%
\bibitem [{\citenamefont {Carrera}\ and\ \citenamefont
  {Giulini}(2006)}]{Carrera:2006im}%
  \BibitemOpen
  \bibfield  {author} {\bibinfo {author} {\bibfnamefont {Matteo}\ \bibnamefont
  {Carrera}}\ and\ \bibinfo {author} {\bibfnamefont {Domenico}\ \bibnamefont
  {Giulini}},\ }\bibfield  {title} {\enquote {\bibinfo {title} {{On the
  influence of the global cosmological expansion on the local dynamics in the
  solar system}},}\ }\href@noop {} {\  (\bibinfo {year} {2006})},\ \Eprint
  {http://arxiv.org/abs/gr-qc/0602098} {arXiv:gr-qc/0602098 [gr-qc]}
  \BibitemShut {NoStop}%
\bibitem [{\citenamefont {Hojman}\ and\ \citenamefont
  {Asenjo}(2013)}]{Hojman:2012me}%
  \BibitemOpen
  \bibfield  {author} {\bibinfo {author} {\bibfnamefont {Sergio~A.}\
  \bibnamefont {Hojman}}\ and\ \bibinfo {author} {\bibfnamefont {Felipe~A.}\
  \bibnamefont {Asenjo}},\ }\bibfield  {title} {\enquote {\bibinfo {title}
  {{Can gravitation accelerate neutrinos?}}}\ }\href {\doibase
  10.1088/0264-9381/30/2/025008} {\bibfield  {journal} {\bibinfo  {journal}
  {Class. Quant. Grav.}\ }\textbf {\bibinfo {volume} {30}},\ \bibinfo {pages}
  {025008} (\bibinfo {year} {2013})},\ \Eprint {http://arxiv.org/abs/1203.5008}
  {arXiv:1203.5008 [physics.gen-ph]} \BibitemShut {NoStop}%
\bibitem [{\citenamefont {Verhaaren}\ and\ \citenamefont
  {Hirschmann}(2010)}]{Verhaaren:2009md}%
  \BibitemOpen
  \bibfield  {author} {\bibinfo {author} {\bibfnamefont {Chris}\ \bibnamefont
  {Verhaaren}}\ and\ \bibinfo {author} {\bibfnamefont {Eric~W.}\ \bibnamefont
  {Hirschmann}},\ }\bibfield  {title} {\enquote {\bibinfo {title} {{Chaotic
  orbits for spinning particles in Schwarzschild spacetime}},}\ }\href
  {\doibase 10.1103/PhysRevD.81.124034} {\bibfield  {journal} {\bibinfo
  {journal} {Phys. Rev.}\ }\textbf {\bibinfo {volume} {D81}},\ \bibinfo {pages}
  {124034} (\bibinfo {year} {2010})},\ \Eprint {http://arxiv.org/abs/0912.0031}
  {arXiv:0912.0031 [gr-qc]} \BibitemShut {NoStop}%
\bibitem [{\citenamefont {Barack}\ and\ \citenamefont
  {Sago}(2009)}]{Barack:2009ey}%
  \BibitemOpen
  \bibfield  {author} {\bibinfo {author} {\bibfnamefont {Leor}\ \bibnamefont
  {Barack}}\ and\ \bibinfo {author} {\bibfnamefont {Norichika}\ \bibnamefont
  {Sago}},\ }\bibfield  {title} {\enquote {\bibinfo {title} {{Gravitational
  self-force correction to the innermost stable circular orbit of a
  Schwarzschild black hole}},}\ }\href {\doibase
  10.1103/PhysRevLett.102.191101} {\bibfield  {journal} {\bibinfo  {journal}
  {Phys. Rev. Lett.}\ }\textbf {\bibinfo {volume} {102}},\ \bibinfo {pages}
  {191101} (\bibinfo {year} {2009})},\ \Eprint {http://arxiv.org/abs/0902.0573}
  {arXiv:0902.0573 [gr-qc]} \BibitemShut {NoStop}%
\bibitem [{\citenamefont {Jefremov}\ \emph {et~al.}(2015)\citenamefont
  {Jefremov}, \citenamefont {Tsupko},\ and\ \citenamefont
  {Bisnovatyi-Kogan}}]{Jefremov:2015gza}%
  \BibitemOpen
  \bibfield  {author} {\bibinfo {author} {\bibfnamefont {Paul~I.}\ \bibnamefont
  {Jefremov}}, \bibinfo {author} {\bibfnamefont {Oleg~{\relax Yu}.}\
  \bibnamefont {Tsupko}}, \ and\ \bibinfo {author} {\bibfnamefont {Gennady~S.}\
  \bibnamefont {Bisnovatyi-Kogan}},\ }\bibfield  {title} {\enquote {\bibinfo
  {title} {{Innermost stable circular orbits of spinning test particles in
  Schwarzschild and Kerr space-times}},}\ }\href {\doibase
  10.1103/PhysRevD.91.124030} {\bibfield  {journal} {\bibinfo  {journal} {Phys.
  Rev.}\ }\textbf {\bibinfo {volume} {D91}},\ \bibinfo {pages} {124030}
  (\bibinfo {year} {2015})},\ \Eprint {http://arxiv.org/abs/1503.07060}
  {arXiv:1503.07060 [gr-qc]} \BibitemShut {NoStop}%
\bibitem [{\citenamefont {Abramowicz}\ \emph {et~al.}(2010)\citenamefont
  {Abramowicz}, \citenamefont {Jaroszynski}, \citenamefont {Kato},
  \citenamefont {Lasota}, \citenamefont {Rozanska},\ and\ \citenamefont
  {Sadowski}}]{Abramowicz:2010nk}%
  \BibitemOpen
  \bibfield  {author} {\bibinfo {author} {\bibfnamefont {Marek~A.}\
  \bibnamefont {Abramowicz}}, \bibinfo {author} {\bibfnamefont {Michal}\
  \bibnamefont {Jaroszynski}}, \bibinfo {author} {\bibfnamefont {Shoji}\
  \bibnamefont {Kato}}, \bibinfo {author} {\bibfnamefont {Jean-Pierre}\
  \bibnamefont {Lasota}}, \bibinfo {author} {\bibfnamefont {Agata}\
  \bibnamefont {Rozanska}}, \ and\ \bibinfo {author} {\bibfnamefont
  {Aleksander}\ \bibnamefont {Sadowski}},\ }\bibfield  {title} {\enquote
  {\bibinfo {title} {{Leaving the ISCO: the inner edge of a black-hole
  accretion disk at various luminosities}},}\ }\href {\doibase
  10.1051/0004-6361/201014467} {\bibfield  {journal} {\bibinfo  {journal}
  {Astron. Astrophys.}\ }\textbf {\bibinfo {volume} {521}},\ \bibinfo {pages}
  {A15} (\bibinfo {year} {2010})},\ \Eprint {http://arxiv.org/abs/1003.3887}
  {arXiv:1003.3887 [astro-ph.HE]} \BibitemShut {NoStop}%
\bibitem [{\citenamefont {Giblin}\ \emph {et~al.}(2004)\citenamefont {Giblin},
  \citenamefont {Marolf},\ and\ \citenamefont {Garvey}}]{Giblin:2003xj}%
  \BibitemOpen
  \bibfield  {author} {\bibinfo {author} {\bibfnamefont {John~T.}\ \bibnamefont
  {Giblin}, \bibfnamefont {Jr.}}, \bibinfo {author} {\bibfnamefont {Donald}\
  \bibnamefont {Marolf}}, \ and\ \bibinfo {author} {\bibfnamefont {Robert}\
  \bibnamefont {Garvey}},\ }\bibfield  {title} {\enquote {\bibinfo {title}
  {{Space-time embedding diagrams for spherically symmetric black holes}},}\
  }\href {\doibase 10.1023/B:GERG.0000006695.17232.2e} {\bibfield  {journal}
  {\bibinfo  {journal} {Gen. Rel. Grav.}\ }\textbf {\bibinfo {volume} {36}},\
  \bibinfo {pages} {83--99} (\bibinfo {year} {2004})},\ \Eprint
  {http://arxiv.org/abs/gr-qc/0305102} {arXiv:gr-qc/0305102 [gr-qc]}
  \BibitemShut {NoStop}%
\bibitem [{\citenamefont {de~la Cruz-Dombriz}\ \emph
  {et~al.}(2009)\citenamefont {de~la Cruz-Dombriz}, \citenamefont {Dobado},\
  and\ \citenamefont {Maroto}}]{delaCruzDombriz:2009et}%
  \BibitemOpen
  \bibfield  {author} {\bibinfo {author} {\bibfnamefont {A.}~\bibnamefont
  {de~la Cruz-Dombriz}}, \bibinfo {author} {\bibfnamefont {A.}~\bibnamefont
  {Dobado}}, \ and\ \bibinfo {author} {\bibfnamefont {A.~L.}\ \bibnamefont
  {Maroto}},\ }\bibfield  {title} {\enquote {\bibinfo {title} {{Black Holes in
  f(R) theories}},}\ }\href {\doibase 10.1103/physrevd.83.029903,
  10.1103/PhysRevD.80.124011} {\bibfield  {journal} {\bibinfo  {journal} {Phys.
  Rev.}\ }\textbf {\bibinfo {volume} {D80}},\ \bibinfo {pages} {124011}
  (\bibinfo {year} {2009})},\ \bibinfo {note} {[Erratum: Phys.
  Rev.D83,029903(2011)]},\ \Eprint {http://arxiv.org/abs/0907.3872}
  {arXiv:0907.3872 [gr-qc]} \BibitemShut {NoStop}%
\end{thebibliography}%

\end{document}